\documentclass[prd,preprint,nofootinbib,superscriptaddress
,secnumarabic,showpacs,showkeys%
,amssymb, amsmath,mathrsfs,nobibnotes,aps,11pt]{revtex4}
\usepackage{graphicx,amsmath,amsfonts,amssymb,dcolumn,epsfig,bm,color}
\usepackage{graphicx, multirow}
\usepackage{dcolumn}
\usepackage{bm}
\usepackage{capt-of}
\usepackage{enumerate}
\usepackage{epsfig} 
\usepackage{wasysym} 
\usepackage{graphicx} 
\usepackage{amsfonts} 
\usepackage{amsbsy} 
\usepackage{amscd} 
\usepackage{hyperref}
\usepackage{pstricks} 
\usepackage{multirow}
\usepackage{ulem}
\usepackage{slashbox}
\usepackage{url}

\makeatletter

\newcommand{\fmslash}[2][0mu]{%
  \mathchoice
    {\fmsl@sh\displaystyle{#1}{#2}}%
    {\fmsl@sh\textstyle{#1}{#2}}%
    {\fmsl@sh\scriptstyle{#1}{#2}}%
    {\fmsl@sh\scriptscriptstyle{#1}{#2}}}
\newcommand{\fmsl@sh}[3]{%
  \m@th\ooalign{$\hfil#1\mkern#2/\hfil$\crcr$#1#3$}}
\makeatother

\newcommand{\lsim}{{\;\raise0.3ex\hbox{$<$\kern-0.75em\raise-1.1ex\hbox{$\sim$}}\;}}
\newcommand{\gsim}{{\;\raise0.3ex\hbox{$>$\kern-0.75em\raise-1.1ex\hbox{$\sim$}}\;}}

\newcommand{\MET}{{{\fmslash E}_T}}
\def\gs{\mathrel{
   \rlap{\raise 0.511ex \hbox{$>$}}{\lower 0.511ex \hbox{$\sim$}}}}
\def\ls{\mathrel{
   \rlap{\raise 0.511ex \hbox{$<$}}{\lower 0.511ex \hbox{$\sim$}}}}

\begin{document}
  \title{Constraints on a seesaw model leading to Quasi-Degenerate neutrinos \\ and signatures at the LHC} 
  
\author{Gulab Bambhaniya}
\affiliation{Physical Research Laboratory, Ahmedabad-380009, Gujarat, India}
\author{Subrata Khan}
\affiliation{Physical Research Laboratory, Ahmedabad-380009, Gujarat, India}
\author{Partha Konar}
\affiliation{Physical Research Laboratory, Ahmedabad-380009, Gujarat, India}
\author{Tanmoy Mondal}
\email{gulab, subrata, konar, tanmoym@prl.res.in}
\affiliation{Physical Research Laboratory, Ahmedabad-380009, Gujarat, India}
\affiliation{Indian Institute of Technology, Gandhinagar-382424, Gujarat, India.}

\begin{abstract}
We consider a variant of TeV scale seesaw models in which three additional heavy right handed neutrinos are 
added to the standard model  to generate the quasi-degenerate light neutrinos.  
This model is theoretically interesting since it can be fully rebuilt from the experimental data 
of neutrino oscillations except for an unknown factor in the Dirac Yukawa coupling.
We study the constrains on this coupling coming from meta-stability of electro-weak vacuum.
Even stronger bound comes from the lepton flavor violating decays on this model, 
especially in a heavy neutrino mass scenario which is within the collider reach.
Bestowed with these constrained parameters, we explore the production and discovery potential
coming from these heavy neutrinos at the $14$~TeV run of Large Hadron Collider.
Signatures with tri-lepton final state together with backgrounds are considered in a realistic 
simulation.
\end{abstract}



% 12.60.-i  : Models beyond the standard model
% 13.85.Qk   Inclusive production with identified leptons, photons, or other nonhadronic particles
% 14.60.St : Non-standard-model neutrinos, right-handed neutrinos, etc.
% 13.35.Hb : Decays of heavy neutrinos

\maketitle
\newpage
%%%%%%%%%%%%%%%%%%%%%%%%%%%%%%%%%%%%%%%%%%%%%%%%%%%%%%%%%%%%%%%%%%%%%%%%%%%%%%%%%%%%%%%%%%%%%%%%%%%%%%%%%%%%%%%% 
\section{Introduction}
%%%%%%%%%%%%%%%%%%%%%%%%%%%%%%%%%%%%%%%%%%%%%%%%%%%%%%%%%%%%%%%%%%%%%%%%%%%%%%%%%%%%%%%%%%%%%%%%%%%%%%%%%%%%%%%% 
Recent discovery of neutral scalar~\cite{:2012gu,:2012gk} with a mass around 126 GeV 
and gradual confirmation of its Standard Model (SM) Higgs like nature settled the most convincing 
and self consisting model of particle physics. However, several experimental observations along 
with theoretical questions keep high energy physics community unconvinced that we have yet found 
our ultimate theory and complete periodic table of particles. So the quest for a new physics 
beyond the standard model is underway both theoretically and experimentally especially 
with Large Hadron Collider exploring the new horizon of energy and luminosity.

Breakthrough with the Higgs boson also opens up the possibility of exploring new physics by 
studying the stability of the electroweak vacuum~\cite{Casas:1999cd,Gogoladze:2008ak}. For the SM to be the only valid theory, 
vacuum should be stable up to Planck scale $M_P$ ($1.2\times 10^{19}$ GeV) which indicates 
that the Higgs self-coupling must remain positive through Renormalization Group (RG) 
running up to the Planck scale~\cite{Shaposhnikov:2009pv,Holthausen:2011aa,Bezrukov:2012sa,Alekhin:2012py}. 
However, it has been shown~\cite{Degrassi:2012ry} that 
achieving absolute stability within the SM is severely restricted. 
Yet the self coupling is not largely negative near Planck scale which implies 
that the SM vacuum might be metastable~\cite{Isidori:2001bm,Espinosa:2007qp}. This hypothesis can act as a window to 
explore new physics considering that the SM vacuum should not go to instable 
region~\cite{Ellis:2009tp,EliasMiro:2011aa,Degrassi:2012ry}. 
At least it should remain in the metastable region after inclusion of the effect of 
new physics.

Seesaw models those lead to light neutrino masses are 
studied in the context of (meta) stability of the electroweak 
vacuum~\cite{Casas:1999cd,Gogoladze:2008ak,Rodejohann:2012px,Chakrabortty:2012np,Chen:2012faa,Khan:2012zw,Buttazzo:2013uya,
Branchina:2013jra,Branchina:2014usa,Branchina:2014rva}, 
lepton flavor violating (LFV) decay~\cite{Petcov:2005jh,Dinh:2012bp,Abada:2014kba}, neutrino less double beta decay 
($0\nu\beta\beta$) (for a recent review, see~\cite{Rodejohann:2012xd,Bilenky:2014uka}) and new physics signatures of such models at 
present colliders~\cite{Huitu:1996su,Akeroyd:2005gt,Han:2006ip,delAguila:2007em,Han:2007bk,Akeroyd:2007zv,
Bray:2007ru,Perez:2008ha,delAguila:2008cj,Franceschini:2008pz,Xing:2009mm,Atre:2009rg,Melfo:2011nx,Chen:2011de,Eboli:2011ia,
Vanini:2012mla,Das:2012ze,Bambhaniya:2013yca,Dev:2013wba,Aguilar-Saavedra:2013twa,Das:2014jxa,Bambhaniya:2014kga}. 
Seesaw models which consist of extra heavy fields added to the SM predict hierarchical light neutrino mass 
spectrum (such as, normal hierarchy and inverted hierarchy) as well as degenerate light neutrino mass 
spectrum~\cite{Minkowski:1977sc,Yanagida:1979as,GellMann:1980vs,Glashow:1979nm,Mohapatra:1979ia,Schechter:1980gr}. 
With recent results from Planck data~\cite{Ade:2013zuv}, degenerate mass spectrum becomes severely restricted, 
although quasi-degenerate (QD) mass spectrum
~\cite{Minkowski:1977sc,Yanagida:1979as,GellMann:1980vs,Glashow:1979nm,Mohapatra:1979ia,Schechter:1980gr} 
is not fully ruled out. It is worthwhile to study QD models in the light of new constraints coming from vacuum
(meta)stability and lepton flavor violation (LFV), also to investigate the possibility of observing signatures
of this model at the upcoming 14 TeV LHC.

In this paper we consider a variant of TeV scale seesaw models consists of three heavy neutrinos along with the SM, which leads to 
quasi-degenerate light neutrino mass spectrum.  We explore the constraints on the parameters (neutrino Yukawa matrix) 
coming from the metastability bound. The neutrino Yukawa matrix is constrained significantly from the metastability condition while 
having weak dependence on right handed heavy neutrino mass~\cite{Khan:2012zw}. 
The experimental uncertainties from top quark mass, strong coupling constant, and particularly 
those from the neutrino data permit a notable window in the constrained value of neutrino Yukawa coupling. 
The allowed parameter space has been restricted further by combining it with the bound coming from 
lepton flavor violating (LFV) decay process such as $\mu \to e\,\gamma$. 
However, LFV bound strongly depends on the unknown phases of the Pontecorvo-Maki-Nakagawa-Sakata matrix ($U_{PMNS}$). 
Considering the best fit values of oscillation parameters, one
would find the bulk of the parameter space (unknown phases), depending upon the choice
of these parameters the LFV constraint can be more restrictive compared to
metastability bound up to $\sim$ TeV.

Once we found the constrained parameters in this model where neutrino Yukawa matrix is fully 
reconstructible with the present oscillation data, we  study the collider signatures of the 
heavy neutrinos at 14 TeV LHC. Heavy neutrinos can be produced dominantly through $s$-channel 
production process associated with lepton which subsequently produce tri-lepton signal along 
with missing transverse energy coming from non-detection of light neutrino.
We have considered the leading order production and performed the particle level realistic 
simulation to estimate this signal using {\tt MadGraph} and {\tt PYTHIA}. 
Besides $s$-channel process, heavy neutrino can also be produced through vector boson 
fusion (VBF) process, where weak gauge bosons originating from two oppositely moving partons `fuse' 
to produce these heavy neutrinos. In the VBF production channel, the final new physics signal is 
accompanied by two forward tagged jets. Since there is no color connection between the two forward 
tagged jets, the central region is devoid of any color activity. This significantly lowers the 
background making weak signals more prominent. These features were exploited not only in the Higgs search 
(see,~\cite{Rainwater:1999gg} and references therein), but also proposed as an avenue to explore 
new physics~\cite{Datta:2001cy, Datta:2001hv,Choudhury:2003hq,Cho:2006sx} at the LHC.
However, in our case we found that the VBF production cross section of heavy neutrino  
is too low to provide  any conclusive signature in the proposed luminosity. 

Organization of the paper goes as follows: Sec.~\ref{sec:model} contains a brief description 
of the model leading to the quasi-degenerate light neutrinos. Vacuum metastability and LFV 
bounds are discussed in Sec.~\ref{sec:meta_bound} and Sec.~\ref{sec:lfv_bound}  respectively. 
We also briefly discuss neutrino less double beta decay in this model in Sec.~\ref{sec:0nubetabeta}. 
Thereafter we proceed for collider search strategy by discussing the heavy neutrino production 
channels at the LHC and its decay in Sec.~\ref{sec:phenomenology}. Detailed simulation, 
event selection criteria together with expected signal and background results are presented in 
sec.~\ref{sec:simulation_result} followed by discovery potential in Sec.~\ref{sec:discovery_pot}.  
Finally we summarize and conclude in Sec.~\ref{sec:conclusion}.

%%%%%%%%%%%%%%%%%%%%%%%%%%%%%%%%%%%%%%%%%%%%%%%%%%%%%%%%%%%%%%%%%%%%%%%%%%%%%%%%%%%%%%%%%%%%%%%%%%%%%%%%%%%%%%%% 
\section{The model}
\label{sec:model}
%%%%%%%%%%%%%%%%%%%%%%%%%%%%%%%%%%%%%%%%%%%%%%%%%%%%%%%%%%%%%%%%%%%%%%%%%%%%%%%%%%%%%%%%%%%%%%%%%%%%%%%%%%%%%%%%
We extend the Standard Model (SM) particle spectra by adding three heavy right 
handed neutrinos having mass at TeV scale. 
The additional part of the Lagrangian is given by 
\begin{eqnarray} 
\mathcal{L}_{ext} = -\,\tilde{\phi}^{\dagger} \overline{N}_R Y_{\nu} l_L 
- \frac{1}{2} \overline{N}_R \,M\, N_R^c + \textrm{H.c.}\,, 
\label{eq:yukawa_NR} 
\end{eqnarray}
where $l_L$ is the left handed lepton doublet, 
$\phi$ is the SM Higgs doublet and $\tilde{\phi}$ 
is given by $\tilde{\phi} = i \sigma^2 \phi^{*}$. The right handed singlet heavy 
neutrino field is denoted by $N_R$. $(Y_{\nu})_{ji}$ are the elements of the Dirac Yukawa coupling matrix 
of dimension $(3 \times 3)$ in the present model with first(second) index is assigned for heavy(light) 
neutrinos. After spontaneous symmetry breaking the Higgs field acquire vacuum expectation value $v$, consequently the light 
neutrino mass matrix is given by
\begin{eqnarray}
m_{\nu} = m_D^T\, M^{-1} m_D.
\label{eq:seesaw1} 
\end{eqnarray}
Where Dirac mass term is given by $m_D = Y_{\nu}v/\sqrt{2}$. 
Using the parameterization based on Casas and Ibarra~\cite{Casas:2001sr}, 
 texture of the Yukawa coupling matrix $Y_{\nu}$ can be expressed as{\footnote{For two heavy neutrino case, 
 the parameterization has been studied by Ibarra et.al. \cite{Ibarra:2003up}.}}
\begin{eqnarray}
Y_{\nu} = \frac{\sqrt{2}}{v} \sqrt{M^d} \,R \sqrt{m^d_\nu} \,U_{\tt PMNS}^{\dagger},
{\label{eq:cas-ibr}}
\end{eqnarray}
where $M^d$ and $m^d_{\nu}$ are the heavy and light neutrino mass matrices 
respectively in their diagonal basis\footnote{In the present work we have taken $M$ to be diagonal which 
implies $M$ and $M^d$ are equivalent.}. $U_{\tt PMNS}$ is the light neutrino mixing matrix, given by 
\begin{eqnarray}
U_{\tt PMNS}  =  
 \begin{pmatrix}
 c_{12}\,c_{13} & s_{12}\,c_{13} & s_{13}\,e^{-i \delta} \\ 
-\,c_{23}\,s_{12} - s_{23}\, s_{13}\, c_{12}\, e^{i \delta} & c_{23}\, c_{12}-s_{23}\, s_{13}\, s_{12}\,e^{i \delta} & s_{23}\,c_{13} \\ 
s_{23}\, s_{12} - \, c_{23}\, s_{13}\, c_{12}\, e^{i \delta} & -\,s_{23}\, c_{12}-c_{23}\, s_{13}\, s_{12}\,e^{i \delta} & c_{23}\, c_{13}
 \end{pmatrix} P \, ,
\label{eq:upmns_param}
\end{eqnarray}
with $c_{mn} = \cos \theta_{mn}$, $s_{mn} = \sin \theta_{mn}$ and $\delta$ is the Dirac CP phase. $P$ is 
the Majorana phase matrix, expressed as $P = {\rm diag}( e^{-i \alpha_1/2}, e^{-i \alpha_2/2},1)$.
For this parameterization of $Y_{\nu}$, clearly measurable parameters from the low energy 
neutrino experiments enters through $m_{\nu}^d$ and $U_{\tt PMNS}$. Whereas all unknown parameters are 
originated from $M^d$ as well as from complex orthogonal matrix $R$. For simplicity $M^d$ has been 
approximated with a single parameter of heavy neutrino mass. 
Elements of the matrix $R$  are completely arbitrary and can be very large which eventually elevate  
the Yukawa couplings (cf. Eq.~\ref{eq:cas-ibr}) to $\mathcal{O}(1)$. On the other hand, owing to the relation $R R^T = I$, 
this arbitrary elements do not effect in the determination of $m_\nu$ as in Eq.~\ref{eq:seesaw1}. In other words, 
the matrix $R$ acts like a fine tuning parameter which helps to generate sufficiently large Yukawa along with TeV scale $M_R$.

\begin{figure} [t!]
\begin{center}
\includegraphics[width=8.0cm,height=5.5cm]{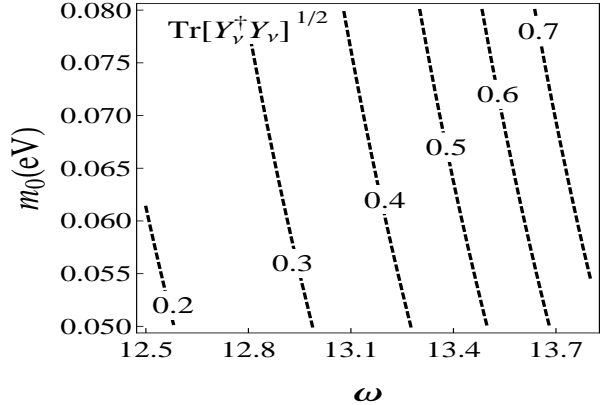} %\\
\caption{Parametric plot of $(\text{Tr}[Y_\nu^\dag Y_\nu])^{1/2}$  with $\omega$ and common 
light neutrino mass scale $m_0$. Heavy neutrino mass fixed at $100$ GeV. 
The numbers in the plot indicates the corresponding values for the 
different set of parameters $\omega$ and $m_0$. }
\label{fig:param-ynudagynu_omwga-m0}
\end{center}
\end{figure}

Orthogonality ensures that the matrix $R$  can be written as,
\begin{eqnarray}\label{eq:rmatrix}
 R = O\, e^{i A},
\end{eqnarray}
where $O$ and $A$ are real orthogonal\footnote{Satisfying $det[O]=det[R]$.} and real antisymmetric matrices respectively. 
For nearly degenerate light neutrinos one can absorb $O$ in the $U_{\tt PMNS}$~\cite{Pascoli:2003rq}.
General form of the antisymmetric matrix $A$ can be expressed in terms of three unknown parameters
\begin{eqnarray}
A = \begin{pmatrix}
0 & a & b\\
-a & 0 & c \\
-b & -c & 0 
\end{pmatrix},
\label{eq:matrixA}
\end{eqnarray}
with $a,b,c ~ \in ~\mathbb{R}^1$. Expanding and rewriting in terms of a new parameter $\omega=\sqrt{a^2 +b^2 + c^2}$ one would obtain
\begin{eqnarray}
e^{i A} = \mathfrak{1} - \frac{\cosh \omega -1}{\omega^2} \, A^2 + i \, \frac{\sinh \omega}{\omega} \, A\,.
\label{eq:exp-A}
\end{eqnarray}
In order to reduce the number of free parameters in our analysis, we choose $a=b=c=\omega/\sqrt{3}$.
Now, we are left with a single unknown parameter $\omega$ (together with single unknown heavy neutrino mass scale $M_R$ as 
diagonal entries of matrix $M^d$) that will be constrained by
imposing the bound of metastability of the electroweak vacuum and non observation of 
LFV decay process.  These constraints would in turn be reflected in terms of norm for Yukawa coupling matrix 
which is extremely crucial in production of the heavy neutrinos and essentially determine the discovery potential at the collider.
Since $Y_\nu$ is a complex square matrix of dimension three, magnitude of which can be 
best represented in terms of the norm of the $Y_\nu$,
\begin{eqnarray}
  \text{Tr}[Y_\nu^\dag Y_\nu] &=& \frac{2\,M_R}{v^2}\,\text{Tr}\left[\sqrt{m_\nu^d}\, 
 R^{\dag} R\,\sqrt{m_\nu^d}\right], \label{eq:ynu-dag-ynu} \\
  &=& \frac{2\,M_R}{v^2}\,m_0\left(1 + 2\,\text{cosh}(\,2\,\omega) \right).
 \label{eq:ynu-dag-ynu_2}
\end{eqnarray}
One can arrive at the much compact expression\footnote{Note that, choice of equal 
$a,b,c$ parameters does not affect this expression. However, unequal parameters would significantly complicate the 
LFV calculation  in Eq.~\ref{eq:mu2e-gamma_2}. Also note that if one of the parameters ($a,b\textrm{ or }c$) is zero, 
then also it is possible to satisfy LFV bound, but it is not the case when two parameters are zero.} 
in terms of the parameter $\omega$, as shown 
in the last equation, assuming an exact degenerate common light neutrino mass scale $m_0$.
For demonstration, contours of constant values of $(\text{Tr}[Y_\nu^\dag Y_\nu])^{1/2}$ is shown in 
Fig.~\ref{fig:param-ynudagynu_omwga-m0} with these parameters.
For our analysis, the common mass scale for light neutrinos is chosen
to be $m_0 \simeq 0.07$ eV, whereas heavy neutrino mass is fixed at $100$ GeV. 
We note that the present allowed light neutrino mass can maximally access the quasi-degenerate range,
and hence the hierarchical neutrino mass can not be neglected completely. One can parameterize  
this effect so that the observed neutrino mass hierarchy can be correctly accommodated within this framework 
of quasi-degenerate neutrinos. We classify them as `normal' and `inverted' hierarchy of masses over the common
mass scale for light neutrinos. As evident from the figure, for a fixed value of $m_0$, different values of  
$(\text{Tr}[Y_\nu^\dag Y_\nu])^{1/2}$ can be obtained by varying $\omega$ accordingly~\cite{Goswami:2013lba}.
To present one example, for this particular choice of degenerate light(heavy) neutrino
mass of $0.07$ eV(100 GeV), the norm $({\rm Tr}[Y^{\dagger}_{\nu} Y_{\nu}])^{1/2} \simeq 0.5$
can be considered for choice of the parameter
\footnote{Note that, for this value of $\omega$, elements of the matrix $e^{iA}$ of Eq.~\ref{eq:rmatrix} 
are of $\mathcal{O}(10^6)$ which enhances the Yukawa coupling matrix as in Eq.~\ref{eq:cas-ibr}.} $\omega=13.4$.

%%%%%%%%%%%%%%%%%%%%%%%%%%%%%%%%%%%%%%%%%%%%%%%%%%%%%%%%%%%%%%%%%%%%%%%%%%%%%
%%%%%%%%%%%%%%%%%%%%%%%%%%%%%%%%%%%%%%%%%%%%%%%%%%%%%%%%%%%%%%%%%%%%%%%%%%%%%
%%%%%%%%%%%%%%%%%%%%%%%%%%%%%%%%%%%%%%%%%%%%%%%%%%%%%%%%%%%%%%%%%%%%%%%%%%%%%%%%%%%%%%%%%%%%%%%%%%%%%%%%%%%%%%%%
\section{Metastability bound}\label{sec:meta_bound}
%%%%%%%%%%%%%%%%%%%%%%%%%%%%%%%%%%%%%%%%%%%%%%%%%%%%%%%%%%%%%%%%%%%%%%%%%%%%%%%%%%%%%%%%%%%%%%%%%%%%%%%%%%%%%%%%
\noindent The SM potential at tree level is given as 
\begin{eqnarray} 
\mathcal{V}(\phi) = \lambda \left(\phi^{\dagger} \phi\right)^2 - m^2 \,\phi^{\dagger} \phi\;.
\end{eqnarray} 
The physical Higgs mass, in the above convention, is defined as $m_h^2 = 2\lambda\,v^2$.  
The Renormalization Group equation (RGE) of $\lambda$ can be expressed up to $i^{\rm th}$ 
loop as 
\begin{eqnarray} 
\frac{d\lambda}{d\,{\rm ln}\mu} = \sum_{i} \frac{\beta_\lambda^{(i)}}{(16 \pi^2)^i}\,, 
\end{eqnarray} 
where $\mu$ is the renormalization scale. The $\beta$ function for one loop is given as, 
\begin{eqnarray} 
\beta_\lambda^{(1)} = 24\, \lambda^2 - \left(\frac{9}{5}\,g_1^2 + 9\,g_2^2\right) \lambda 
+ \frac{27}{200}\,g_1^4 + \frac{9}{20}\,g_1^2\,g_2^2 + \frac{9}{8}\,g_2^2 
+ 4 T \lambda - 2 Y\;,
\label{eq:betalam1} 
\end{eqnarray}
where, 
\begin{eqnarray} 
T & = & {\mathrm {Tr}} \left[3 \,{Y_u}^\dag Y_u + 3\, {Y_d}^\dag Y_d + {Y_l}^\dag Y_l 
+ Y_\nu^\dag Y_\nu \right]\,, \\    
Y & = & {\mathrm{Tr}} \left[3 (Y_u^\dag Y_u)^2 + 3 (Y_d^\dag Y_d)^2 
+ (Y_l^\dag Y_l)^2 + (Y_\nu^\dag Y_\nu)^2\right]  
\end{eqnarray} 
and $g_i$'s are the gauge coupling constants. Grand Unified Theory (GUT) 
modification for the $U(1)$ gauge coupling has been incorporated. $Y_u$, $Y_d$ and $Y_l$ 
denote the Yukawa coupling matrices for the up type quark, down type quark and charged lepton respectively. 
Expectedly, dominant contribution comes from the top Yukawa (up type quark) running and one 
loop $\beta$ function is governed by the following equation:
\begin{eqnarray} 
\beta_{Y_u}^{(1)} = Y_u \left[\frac{3}{2} {Y_u}^\dagger Y_u +
\frac{3}{2} {Y_d}^\dagger Y_d  + T - \left(
\frac{17}{20} g_1^2 + \frac{9}{4} g_2^2 + {8} g_3^2\right)\right].
\label{eq:betayu-1loop} 
\end{eqnarray}
Three loop RGE for Higgs self coupling ($\lambda$), the top Yukawa and the gauge couplings has been 
used in the numerical analysis~\cite{Einhorn:1992um,Luo:2002ey,Machacek:1983tz,Machacek:1983fi,
Machacek:1984zw,Antusch:2002rr,Mihaila:2012fm,Chetyrkin:2012rz}. Matching corrections for top Yukawa has been taken 
up to three loop QCD~\cite{Melnikov:2000qh}, one loop electroweak~\cite{Hempfling:1994ar,Schrempp:1996fb} 
and $\mathcal{O}(\alpha\alpha_s)$~\cite{Bezrukov:2012sa,Jegerlehner:2003py} while for Higgs self 
coupling, it has been taken up to two loop~\cite{Degrassi:2012ry,Sirlin:1985ux}. 
The Higgs self coupling also receives additional contribution from the higher order corrections 
of the effective potential. The loop corrected\footnote{We incorporated two loop correction due to the SM and 
one loop correction due to neutrino Yukawa couplings.} effective self coupling denoted by $\tilde{\lambda}$, 
is given by~\cite{Casas:1994qy,Casas:1996aq,Khan:2012zw},
\begin{eqnarray} 
\tilde{\lambda} &=& \lambda - \frac{1}{32\,\pi^2} \left[\frac{3}{8} \left(g_1^2 + g_2^2 \right)^2 \left(\frac{1}{3} - 
{\mathrm{ln}}\frac{\left(g_1^2 + g_2^2 \right)}{4}\right) + 6\,y_t^4 \left({\mathrm{ln}}\,\frac{y_t^2}{2} - 1 \right) 
 + \frac{3}{4}\,g_2^4 \left(\frac{1}{3} - {\mathrm{ln}}\,\frac{g_2^2}{4}\right) \right. \nonumber \\
& & \left . +\, ([Y_\nu^\dag Y_\nu]_{ii})^2
\left({\mathrm{ln}}\,\frac{[Y_\nu^\dag Y_\nu]_{ii}}{2} - 1 \right)
+ ([Y_\nu \, Y_\nu^{\dag}]_{jj})^2 
\left({\mathrm{ln}}\,\frac{[Y_\nu \, Y_\nu^{\dag}]_{jj}}{2} - 1\right) \right] 
+ \frac{Y_t^4}{\left(16\pi^2\right)^2}\times \nonumber \\
& & \left[g_3^2\left\{24\left({\mathrm{ln}}\,\frac{Y_t^2}{2}\right)^2 
- 64\,{\mathrm{ln}}\,\frac{Y_t^2}{2} + 72 \right\} - \frac{3}{2}\,Y_t^2\left\{3\left({\mathrm{ln}}\,\frac{Y_t^2}{2}\right)^2 
- 16\,{\mathrm{ln}}\,\frac{Y_t^2}{2} + 23 + \frac{\pi^2}{3} \right\}\right], 
\label{lambdatilda2}
\end{eqnarray} 
where $i\,,j$ denote the number of generation of light and heavy neutrinos respectively. 
The absolute stability of the electro weak vacuum implies $\tilde{\lambda}\ge 0$ upto 
Planck scale. However as shown in~\cite{Degrassi:2012ry}, the absolute stability is 
highly restrictive. In this light we shall consider metastability {\it i.e.} 
transition time from a metastable vacuum towards instability should be greater 
than the age of the universe. In other words the transition probability through 
quantum tunneling should be less than unity. 

The tunneling probability within the semi-classical approximation is given by (at 
zero temperature)~\cite{Coleman:1977py,Callan:1977pt,Isidori:2001bm,Espinosa:2007qp}
\begin{eqnarray} 
p=\underset{\mu<\Lambda}{\text{max}}~V_U \, \mu^4 \,\,\text{exp}\left(-\frac{8\pi^2}{3|\lambda(\mu)|}\right),
\end{eqnarray} 
where $\Lambda$ is the cutoff scale and $V_U$ is volume of the past light-cone, 
taken as $\tau^4$. Here $\tau$ is the age of the universe taken from Planck data 
as $\tau=4.35\times 10^{17}$ sec~\cite{Ade:2013sjv}. For the vacuum to be metastable, 
one should have $p<1$ which can be recast in terms of a lower bound on $\lambda$, as given below
\begin{eqnarray} 
\left|\lambda\right|<\lambda_{\text{meta}}^{\text{max}}=\frac{8\pi^2}{3}\frac{1}{4\,\text{ln}\left(\tau\mu\right)} .
\label{lambda_meta}
\end{eqnarray}
The above equation can be utilized to put an upper bound on $\text{Tr}[Y_\nu^\dag Y_\nu]$ from the running 
of $\lambda$ as a function of the heavy neutrino mass $M_R$. 
This has been displayed in Fig.~\ref{fig:meta-lfv_QD} as horizontal slanting lines 
corresponding to different choices of the top mass and strong coupling. Now, the region below this line is
consistent with the metastability bound.

%%%%%%%%%%%%%%%%%%%%%%%%%%%%%%%%%%%%%%%%%%%%%%%%%%%%%%%%%%%%%%%%%%%%%%%%%%%%%%%%%%%%%%%%%%%%%%%%%%%%%%%%%%%%%%%%
\section{Lepton Flavor Violation bound}\label{sec:lfv_bound}
%%%%%%%%%%%%%%%%%%%%%%%%%%%%%%%%%%%%%%%%%%%%%%%%%%%%%%%%%%%%%%%%%%%%%%%%%%%%%%%%%%%%%%%%%%%%%%%%%%%%%%%%%%%%%%%%
Lepton flavor violating decay processes get significant contribution from the heavy neutrino due 
to its relatively low mass scale compared to the canonical seesaw mechanism.
The experimental upper limit on $\mu\to e\,\gamma$ processes can be translated 
to an upper bound on ${\text{Tr}}[Y_\nu^\dag Y_\nu]$ as a function 
of $M_R$. Branching ratio of $\mu \rightarrow e\gamma$~\cite{Tommasini:1995ii} is given by
\begin{eqnarray}\label{eq:mu2e-gamma}
\text{Br}\left(\mu \rightarrow e\gamma\right) = \frac{3\,\alpha}{8\,\pi}\left|\sum_j V_{ej}V_{j\mu}^{\dag}f(x_j) \right|^2, \
\end{eqnarray}
where dependence of heavy neutrino mass is expressed in terms of dimensionless parameter $x_j =({M_{R_j}^2}/{m_W^2})$ 
in a slowly varying function,
\begin{eqnarray}
f(x)=\frac{x\left(1-6\,x+3\,x^2+2\,x^3-6\,x^2\,\text{ln}\,x\right)}{2\,(1-x)^4}\,.
\end{eqnarray}

%%%%%%%%%%%%%%%%%%%%%%%%%%%%%%%%%%%%%%%%%%%%%%%%%%%%%%%%%%%%%%%%%%%%%%%%%%%%%
\begin{figure} [t!]
\begin{center}
\includegraphics[width=7.5cm,height=6.5cm,angle=0]{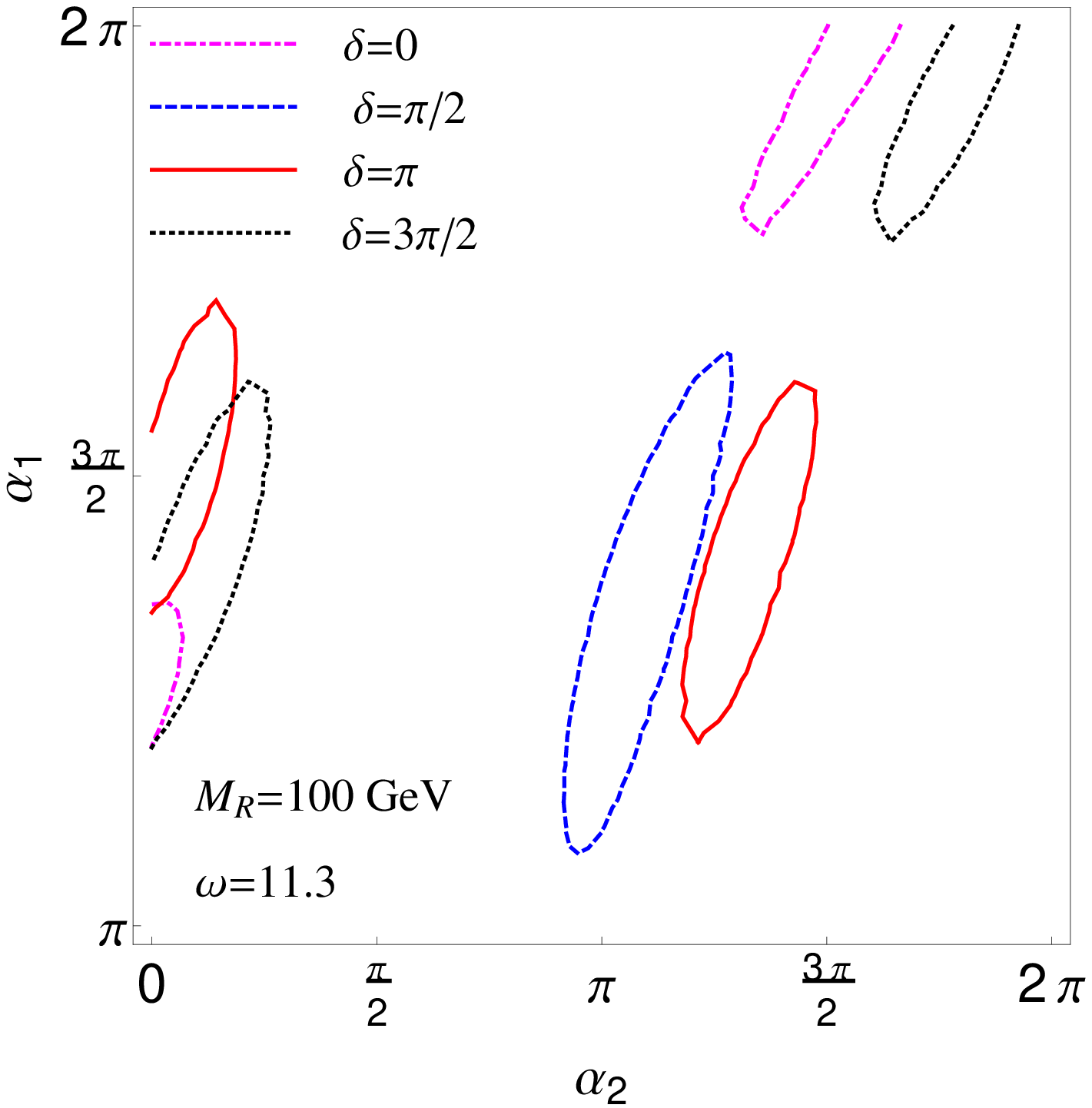} %\\
\includegraphics[width=7.5cm,height=6.5cm,angle=0]{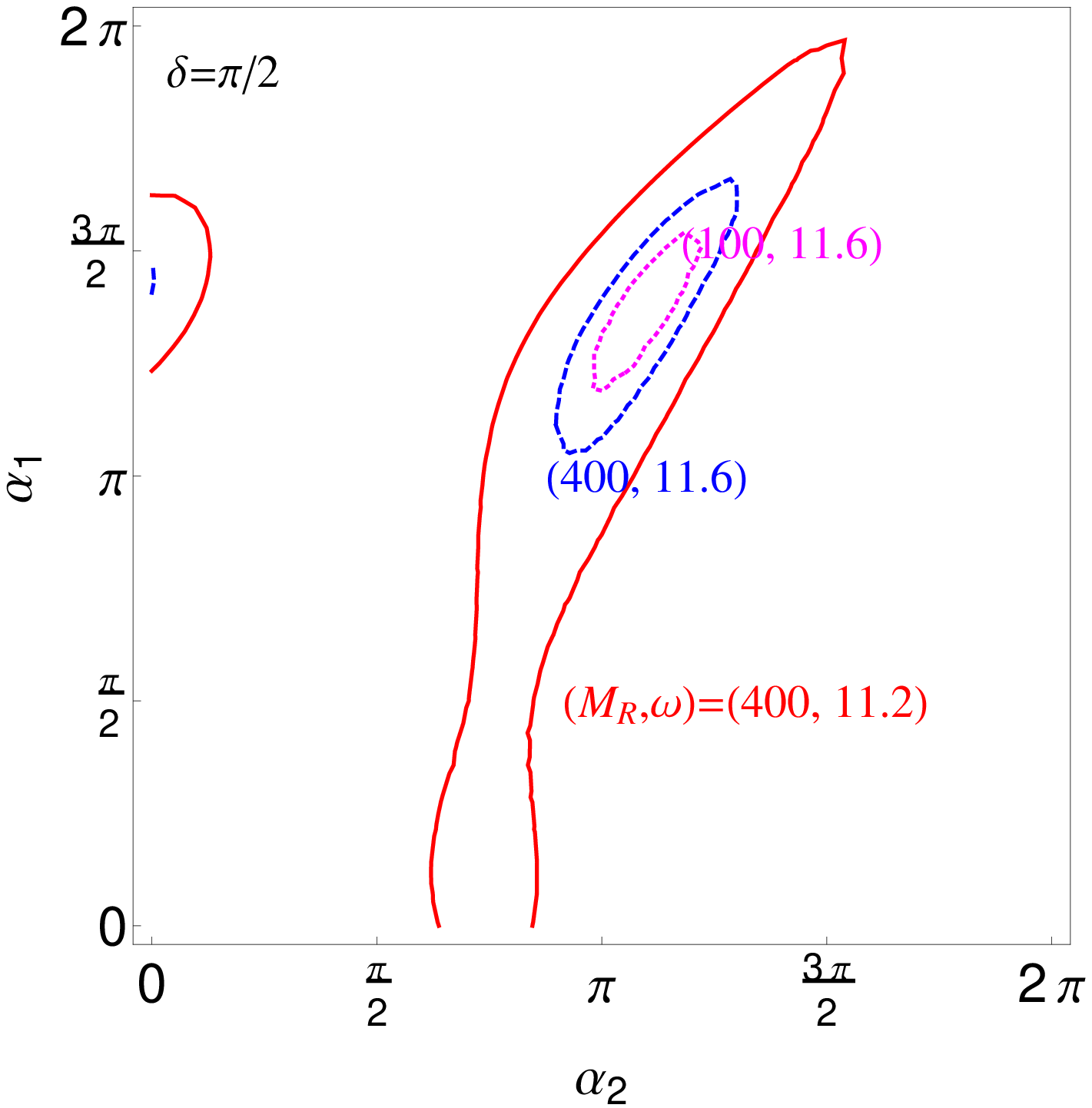}
\caption{(Left panel) Contours of allowed lepton flavor violating regions with 
$\text{Br}\left(\mu \rightarrow e\gamma\right) = 5.7 \times 10^{-13}$ in the parameter plane of 
Majorana phases $\alpha_1$ and $\alpha_2$ with different values of Dirac CP phase $\delta$. 
Considering all the neutrino oscillation parameters and mass differences in the global 
best-fit values, the area within each contours are consistent with the experimental LFV upper bound 
from the decay rate of $\mu\to e\,\gamma$.
(Right panel) demonstrates the variation of these LFV equality
contours for different choices of the heavy neutrino mass $M_R$ and
parameter $\omega$ considering one example ($\delta = \pi/2$) contour from the left panel. As
expected, decreasing the $M_R$ or increasing the $\omega$ would make the contour
narrower, retaining a smaller window for choices of these unknown parameters.}
\label{fig:LFVcontours}
\end{center}
\end{figure}
%%%%%%%%%%%%%%%%%%%%%%%%%%%%%%%%%%%%%%%%%%%%%%%%%%%%%%%%%%%%%%%%%%%%%%%%%%%%%
\begin{figure} [t!]
\begin{center}
\includegraphics[width=8.0cm,height=5.5cm]{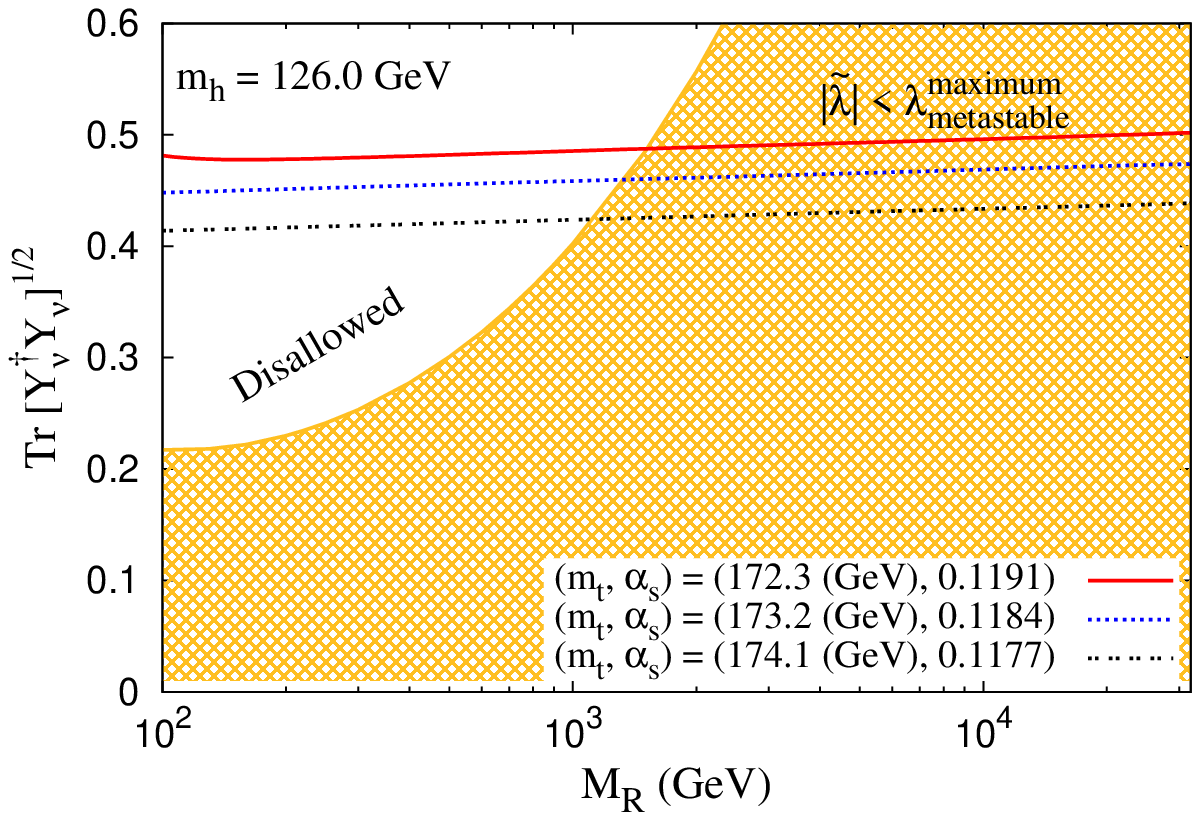} %\\
\caption{Allowed region of the Yukawa norm $\text{Tr}[Y_\nu^\dag Y_\nu]$ as a function of the heavy neutrino mass $M_R$ 
by imposing combined constraints coming from metastability of the electroweak vacuum as well as lepton flavor violating 
decay ($\mu\to e\,\gamma$). Choice of Higgs mass fixed at $m_h=126$ GeV. The horizontal slanting lines represent the upper 
bound on $\text{Tr}[Y_\nu^\dag Y_\nu]$ consistent with the metastability bound, as in Eq.~\ref{lambda_meta}. 
Three lines are due to three different set of values for top mass and strong coupling~\cite{Lancaster:2011wr,Bethke:2009jm}. 
The shaded area below the curved line is allowed from the lepton flavor violating constraint as used in 
Eq.~\ref{brmu2e-gamma}. This is after putting global best-fit values of oscillation parameters 
together with the particular values of unknown phases within their full range as 
tabulated in Table~\ref{table:osc-param-max-min}. The yellow line corresponds to $\omega = 11.9$ and gives us the best
choice for study within the bound of LFV. Hence, the region marked ``Disallowed" is ruled out from LFV
for such choice of $\omega$. }
\label{fig:meta-lfv_QD}
\end{center}
\end{figure}
%%%%%%%%%%%%%%%%%%%%%%%%%%%%%%%%%%%%%%%%%%%%%%%%%%%%%%%%%%%%%%%%%%%%%%%%%%%%%
In our present case,  right handed neutrinos are degenerate, {\it i.e.},  $M_{R_j} = M_R$.
The light-heavy mixing matrix $V$ is obtained through the diagonalization of the 
full neutral lepton mass matrix~\cite{Grimus:2000vj}
\begin{eqnarray}
V = m_D^{\dag} \left(M^{-1}\right)^{\ast} U_R\,,
\label{eq:vln}
\end{eqnarray}
where $U_R$ is a unitary matrix\footnote{$U_R$ is identity matrix in the present scenario as $M$ is diagonal.} 
that diagonalizes $M$. Using Eq.~\ref{eq:cas-ibr} and \ref{eq:vln} with 
Eq.~\ref{eq:mu2e-gamma} one gets,  
\begin{eqnarray}
 \text{Br}\left(\mu \rightarrow e\gamma\right) = \frac{3\,\alpha}{8\,\pi\,M_R^2}\left[f\!\left(\frac{M_R^2}{m_W^2}\right)\right]^2
 \left|U_{\tt PMNS}\sqrt{m_\nu^d}\, R^{\dag} R\,\sqrt{m_\nu^d}\,U_{\tt PMNS}^{\dag} \right|^2
 \label{eq:mu2e-gamma_2}
\end{eqnarray}
and $\text{Tr}[Y_\nu^\dag Y_\nu]$ is given by Eq.~\ref{eq:ynu-dag-ynu_2}. 
From Eq.~\ref{eq:mu2e-gamma_2}, \ref{eq:ynu-dag-ynu} and \ref{eq:ynu-dag-ynu_2} one can see the angular and phase 
dependence of the branching ratio comes from the $U_{\tt PMNS}$, whereas the magnitude of the branching ratio is 
encoded in $\sqrt{m_\nu^d}\,  R^{\dag} R\,\sqrt{m_\nu^d}$ whose modulus is proportional to 
$\text{Tr}[Y_\nu^\dag Y_\nu]$. The analytical expression of $\textrm{~Br}(\mu\to e\gamma)$ is somewhat lengthy and hence omitted here. 
Subjected to the present experimental upper bound on the $\mu\to e\,\gamma$ process~\cite{Adam:2013mnn} 
\begin{eqnarray}
\text{Br}\left(\mu \rightarrow e\gamma\right) \leq 5.7 \times 10^{-13},
\label{brmu2e-gamma} 
\end{eqnarray} 
one would obtain, numerically, an upper bound on $\text{Tr}[Y_\nu^\dag Y_\nu]$ by inverting 
Eq.~\ref{eq:mu2e-gamma_2}.

In a numerical calculation with a very high degree of precision, it is observed that the 
3$\,\sigma$ uncertainty of the oscillation parameters together with all the phases being varied 
in the full range~\cite{Tortola:2012te} would not bound the $\text{Tr}[Y_\nu^\dag Y_\nu]$. 
 Hence  effective bound on $\text{Tr}[Y_\nu^\dag Y_\nu]$ is coming from vacuum metastability only. 
To probe this a little further, in Fig.~\ref{fig:LFVcontours} (left panel) we demonstrate the 
contours of allowed lepton flavor violating regions in the parameter plane of 
Majorana phases $\alpha_1$ and $\alpha_2$ with different values of Dirac CP phase{\footnote{ In the 3$\sigma$ range of oscillation parameters, the Dirac $CP$ phase $\delta$ is allowed in its 
full range (0-2$\pi$).  Also the Majorana phases $\alpha_{1,2}$ are not constrained by oscillation experiments, 
hence are varied in full range (0-2$\pi$). These three phases are considered here as unknown parameters.}} $\delta$. 
Considering all the neutrino oscillation parameters and mass differences at the global best-fit values 
(also listed in Table~\ref{table:osc-param-max-min})~\cite{Tortola:2012te},
the area within each contours are consistent with the experimental LFV upper bound 
from the decay rate of $\mu\to e\,\gamma$. Although not conspicuous from the analytic form of 
multi-parameter expression from Eqn.~\ref{eq:mu2e-gamma_2}, one can evaluate that the suitable 
and precise choice of $\delta$ and $\alpha$ parameters within such contours can indeed evade the 
bound. In the right panel of Fig.~\ref{fig:LFVcontours} we demonstrate the variation of these LFV equality
contours for different choice of the heavy neutrino mass $M_R$ and parameter $\omega$ considering once such 
example ($\delta=\pi/2$) contour from the left panel. As expected, decreasing the $M_R$ or increasing $\omega$ would make the 
contour narrower retaining a smaller window for choices of these unknown parameters.

From our discussions above, one can clearly choose a parameter for any phenomenological analysis 
bounded by metastability. 
However, we took an approach to consider a conservative estimates for $\text{Tr}[Y_\nu^\dag Y_\nu]$ satisfying 
both LFV and vacuum metastability bounds. 
To begin with this, we choose a particular set of oscillation parameters such as the 
global best-fit values of oscillation parameters. Now, if one 
examines the particular choices of these unknown phases which would be just enough to satisfy the equality
of Eq.~\ref{brmu2e-gamma}, they are essentially all the points reside over the contours shown in Fig.~\ref{fig:LFVcontours}. 
All the points inside the contours will give BR($\mu\to e\gamma$) $< 5.7\times 10^{-13}$. Since all the 
contours are drawn with a fixed $\omega$ value, the norm  $\text{Tr}[Y_\nu^\dag Y_\nu]$ will be same over all the contours
 shown in Fig.~\ref{fig:LFVcontours}(left panel).

 Dependency of norm  $\text{Tr}[Y_\nu^\dag Y_\nu]$ as a function of heavy neutrino mass is depicted in Fig.~\ref{fig:meta-lfv_QD}.
The upper bound on the norm  is depicted by the golden solid line for $\omega = 11.9$. 
This gives us the best choice to study within the bound of LFV for this particular value of $\omega$. 
The yellow shaded area below the curve are allowed\footnote{This is over the choice of decreasing 
values of $\omega$ parameters.} from the lepton flavor violating constraint as used in Eq.~\ref{brmu2e-gamma}. 
Hence, the region marked ``Disallowed" is strictly ruled out from LFV
for such choice of $\omega$.

For our analysis, we have used the value of $\text{Tr}[Y_\nu^\dag Y_\nu]$ 
allowed from these constraints which reflects as the conservative parameter. 
To get some notion of related neutrino oscillation parameters, 
we list them in Table~\ref{table:osc-param-max-min} which lead to the upper edge
of the yellow shaded region as described in Fig.~\ref{fig:meta-lfv_QD}. 
Note that any other choices of $\omega$ together with this set of angles and phases 
will reside in the region.

%%%%%%%%%%%%%%%%%%%%%%%%%%%%%%%%%%%%%%%%%%%%%%%%%%%%%%%%%%%%%%%%%%%%%%%%%%%%%%%%%%%%%%%%%%%%%%%%%%%%%%%%%%%%%%%%%%%%%%%%%%%%%
\begin{table}[t!]
\begin{tabular}{|c||c|c|c|c|c|c|c|c|}
\hline
Parameters
& $\theta_{12}$ & $\theta_{23}$ & $\theta_{13}$ &$\Delta_{\textrm{sol}}^2$   & $\Delta_{\textrm{atm}}^2$   &$\delta$ & $\alpha_{1}$ & $\alpha_{2}$ \\
&               &               &               &$[10^{-5} \,\textrm{eV}^2]$ & $[10^{-3} \,\textrm{eV}^2]$ &         &              &\\
\hline\hline
~Used value ~&~ $0.19\,\pi~$ & $0.29\,\pi~$ & $0.05\,\pi~$ & 7.62 & 2.55 & $1.37\,\pi~$ & $1.78\,\pi~$ & $1.67\,\pi~$ \\
\hline
\end{tabular}
\caption{Values of oscillation parameters leading to the upper edge of the yellow shaded 
region  in Fig.~\ref{fig:meta-lfv_QD}. We have used global best-fit values of oscillation parameters except the phases.
}
\label{table:osc-param-max-min}
\end{table}
%%%%%%%%%%%%%%%%%%%%%%%%%%%%%%%%%%%%%%%%%%%%%%%%%%%%%%%%%%%%%%%%%%%%%%%%%%%%%%%%%%%%%%%%%%%%%%%%%%%%%%%%%%%%%%%%%%%%%%%%%%%%%
 
%%%%%%%%%%%%%%%%%%%%%%%%%%%%%%%%%%%%%%%%%%%%%%%%%%%%%%%%%%%%%%%%%%%%%%%%%%%%%%%%%%%%%%%%%%%%%%%%%%%%%%%%%%%%%%%%
\section{Neutrino Less Double Beta Decay}\label{sec:0nubetabeta}
%%%%%%%%%%%%%%%%%%%%%%%%%%%%%%%%%%%%%%%%%%%%%%%%%%%%%%%%%%%%%%%%%%%%%%%%%%%%%%%%%%%%%%%%%%%%%%%%%%%%%%%%%%%%%%%%
In this section we briefly discuss the contribution of this particular model towards neutrino less double beta decay 
($0\nu\beta\beta$). The general expression of half-life for $0\nu\beta\beta$ 
in the context of Type-I seesaw is given by~\cite{Mitra:2011qr,Chakrabortty:2012mh} 
\begin{eqnarray}
 T_{\frac{1}{2}}^{-1} = G\, \frac{|\mathcal{M}_\nu|^2}{m_e^2}
 \left|\sum_i \left(U_{\tt PMNS}\right)^2_{e\,i}\,  \left(m_\nu^d\right)_i  
      +\sum_j \langle p^2\rangle \frac{V_{e\,j}^2}{M_{R_j}} 
 \right|^2\,,
 \label{eq:0nubetabeta} 
\end{eqnarray}
where $G=7.93\times10^{-15}$ yr$^{-1}$, $\mathcal{M}_{\nu}$ is the nuclear matrix element due to light 
neutrino exchange and $m_e$ being the electron mass. $\langle p^2\rangle$ in the second term, 
which is due to the contributions from heavy singlet neutrinos, is given by~\cite{Tello:2010am}
\begin{eqnarray}
 \langle p^2\rangle = - \,m_e \,m_p\frac{\mathcal{M}_N}{\mathcal{M}_{\nu}}\,,
 \label{eq:expn_psq}
\end{eqnarray}
which is taken to be
$\langle p^2\rangle = -\, (182 \text{ MeV})^2$~\cite{Mitra:2011qr}. Here $m_p$ is the proton mass and $\mathcal{M}_N$ is 
the nuclear matrix element due to heavy neutrino exchange. 

The first and the second term in Eq.~\ref{eq:0nubetabeta} represent contributions from light and 
heavy neutrinos respectively and thus summed over corresponding number of light(heavy) neutrinos. 
Accordingly with the help of Eq.~\ref{eq:cas-ibr} and \ref{eq:vln}, the second term can be expressed 
as,
\begin{eqnarray}\label{eq:26}
 \frac{\langle p^2\rangle}{M_R^2}\left(U_{\tt PMNS}\sqrt{m_\nu^d}\,R^{\dag}R^{\,\ast}\sqrt{m_\nu^d}\,U_{\tt PMNS}^T 
 \right)_{e\,e} = \frac{\langle p^2\rangle}{M_R^2}\left(U_{\tt PMNS}\right)_{e\,i}^2\,\left(m_\nu^d\right)_i\,.
\end{eqnarray}
Consequently Eq.~\ref{eq:0nubetabeta} becomes
\begin{eqnarray}
 T_{\frac{1}{2}}^{-1} = G\, \frac{|\mathcal{M}_\nu|^2}{m_e^2}\left(1 + \frac{\langle p^2\rangle}{M_R^2}\right)^2 
 \left|\left(U_{\tt PMNS}\right)_{e\,i}^2\,\left(m_\nu^d\right)_i \right|^2\,.
 \label{eq:0nubetabeta_2}
\end{eqnarray}
One can notice that the contribution on $0\nu\beta\beta$ from heavy neutrinos is extremely tiny, {\it e.g.} only $0.001\%$ of 
the light neutrino contribution can come towards the half-life 
of $0\nu\beta\beta$ even for a heavy neutrino mass of $100$ GeV. This contribution is even suppressed as the  mass increased.
Although light neutrino contribution to the neutrino less double beta decay can be sizable and can possibly be explored in the
 future experiments~\cite{Rodejohann:2012xd}, the heavy neutrino contribution in this scenario can be neglected. 
 This outcome is not surprising if one follows from Eq.~\ref{eq:26}. The large values in the matrix $R$, 
 which is essential to obtain large Dirac Yukawa, gets canceled.
 Finally we get very small value of $(VV^T)_{\ell\ell}$ for same sign di-lepton (SSDL) production. 
 In the same ground collider production of SSDL is suppressed and hence not considered 
 although the heavy neutrino is of Majorana type. Interestingly, this is a general consequence of 
 Casas-Ibarra parameterization when the heavy neutrinos are degenerate.  
 At the same time large Yukawa makes the opposite sign di-lepton cross section 
 (which is proportional to $(VV^\dagger)_{\ell\ell}$) sizable. 
 Large SM background in this channel compelled us to consider for tri-lepton signal at the LHC. 
 In the next section, we would explore the production of these heavy neutrinos at the collider and 
 discuss the discovery potential for 14 TeV  large hadron Collider.

%%%%%%%%%%%%%%%%%%%%%%%%%%%%%%%%%%%%%%%%%%%%%%%%%%%%%%%%%%%%%%%%%%%%%%%%%%%%%%%%%%%%%%%%%%%%%%%%%%%%%%%%%%%%%%%%
\section{Signatures at the LHC}\label{sec:phenomenology}
%%%%%%%%%%%%%%%%%%%%%%%%%%%%%%%%%%%%%%%%%%%%%%%%%%%%%%%%%%%%%%%%%%%%%%%%%%%%%%%%%%%%%%%%%%%%%%%%%%%%%%%%%%%%%%%%

\begin{figure}[t!]
\includegraphics[width=5.5cm,height=7.5cm,angle=270]{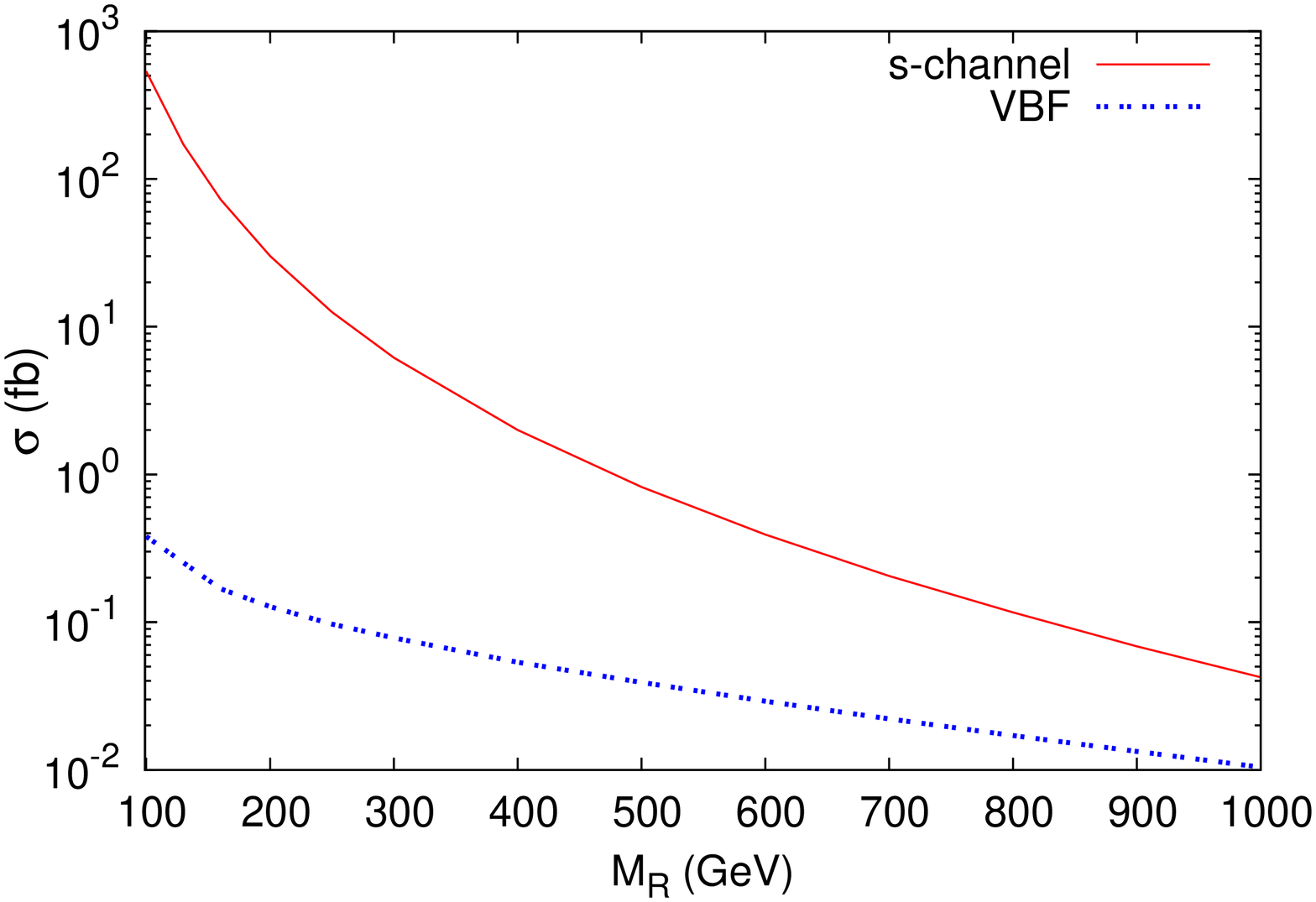}
\includegraphics[width=5.5cm,height=8.0cm,angle=270]{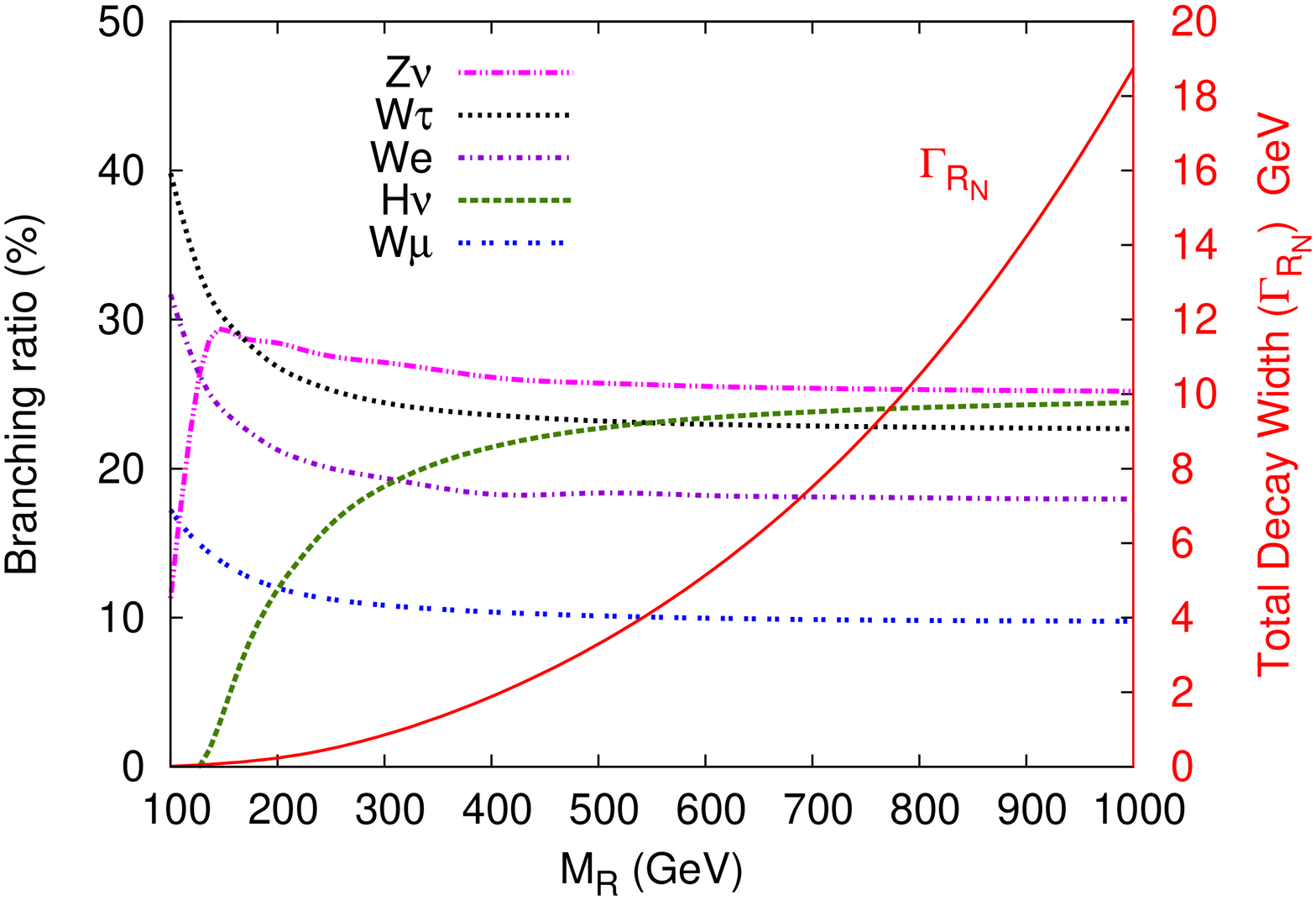}
\caption{(Left panel) Total cross section is plotted for leading order s-channel heavy neutrino production 
(solid line) associated with charged lepton at the 14 TeV LHC. Basic pre-selection cuts
$p_{T\ell} \ge 20$ GeV and $|\eta_\ell| \le 2.5$ are applied and choice of parameters are compatible
with the neutrino oscillation data constrained with vacuum metastability and LFV. 
The dotted line shows the corresponding VBF production cross section, 
where basic VBF cuts were used in addition to the pre-selection cuts. 
(Right panel) Demonstration of the decay branching ratios of the heavy 
 neutrino in different channels as a function of mass. 
 Total decay width is also shown with red-solid line.}
 \label{fig:x-section}
\end{figure}

Heavy neutrinos can be produced dominantly in $s$-channel W-boson exchange at the hadron collider. 
We also explored the corresponding VBF production associated with two forward jets. 
At the leading order calculation, parton level processes producing heavy neutrinos ($N$) at the  mass basis are as follows:
\begin{eqnarray}
 q \bar{q'} &\longrightarrow& {W^{\pm}}^*  \longrightarrow \,l^\pm\, N \;\;\textrm{ (s-channel)}, \nonumber \\ 
 q q' &\longrightarrow& l^\pm\, N \,q\,q''\;\;\; \textrm{ (VBF) },
\end{eqnarray}
where $q$ represents suitable parton and associated leptons are $l \equiv (e,\,\mu,\,\tau)$. 
In Fig.~\ref{fig:x-section}~(left panel) the total cross section for these processes are shown as a 
function of heavy neutrino mass after applying the pre-selection cuts {\it i.e.} $p_{T_l} > 20$ GeV and $|\eta_l|<2.5$. 
The solid (dashed) line is showing leading order production cross section through $s$-channel (VBF) process. 
From the figure it is evident that the VBF cross section is insufficient hence we shall not discuss this 
production mechanism afterwards and concentrate only on $s$-channel process for phenomenological analysis. 

For our simulation we consider the maximum allowed value coming from $\text{Tr}[Y_\nu^\dag Y_\nu]$ 
satisfying combined LFV and meta stability bounds as depicted in Fig.~\ref{fig:meta-lfv_QD} together 
with neutrino oscillation data within their uncertainties.
One can notice that the higher values of Yukawa coupling is permitted from these constraints once we 
move towards higher mass of heavy neutrinos. 
We have used {\tt MadGraph5}~\cite{Alwall:2011uj} to simulate the production and decay of 
heavy neutrinos. Parton distribution function {\tt CTEQ6L1}~\cite{Pumplin:2002vw} has been used  
and the factorization scale is set at heavy neutrino mass.

The heavy neutrino can decay into weak gauge bosons ($W^\pm,Z$) or the Higgs boson ($H$) 
in association with leptons because of mixing between light and heavy neutrinos: 
\begin{equation}
 N \longrightarrow W^\pm\,l^\mp   /   Z \nu_l  /  H \nu_l. 
\end{equation}
Branching ratio of $N$ in these channels are shown in Fig.~\ref{fig:x-section}~(right panel)
with varying heavy neutrino mass $M_R$. In this plot the  red-solid line is showing the total 
decay width ($\Gamma_N$) of heavy neutrino. The figure manifests that $W\tau$ channel  is the  
dominant decay mode for low mass region and saturates at $\sim 22\%$ for $M_R \gtrsim 400$ GeV.
Both $H\nu$ and $Z\nu$ channels saturate  at $\sim 25\%$ in the high mass region leaving approximately 
$18\%(10\%)$ for the $We(W\mu)$ channel. 

% Identifying dominant decay modes for the heavy neutrino, 
One can notice that the decay 
into charged  light leptons ($e,\mu$) associated with onshell $W$ boson can finally produce 
tri-lepton signal with missing transverse momentum at the LHC. This can be vital channel searching
for QD heavy neutrinos at the hadron collider. Since it was shown earlier~\cite{Bambhaniya:2014kga} that the separation 
of these tri-lepton signals into separate flavor states can carry useful informations on the 
hierarchical structures of light neutrinos associated with the model. Hence we would also 
consider flavor allocated cross sections for  signal and the backgrounds.

%%%%%%%%%%%%%%%%%%%%%% TABLE %%%%%%%%%%%%%%%%%%%%%%%%%%%%%%%%%%
\begin{table}[tbh]
\begin{tabular}{|l|l|}
\hline
\multicolumn{2}{|c|}{\Large Selection Criteria} \\
\hline
Lepton identification criteria &  $|\eta_{\ell}| < 2.5$ and ${p_T}_{\ell}  >$  20 GeV\\
\hline
Detector efficiency for leptons & Electron efficiency (for $e^-$ \& $e^+$): 0.7 ($70\%$)\\
&Muon efficiency (for $\mu^-$ \& $\mu^+$): 0.9 ($90\%$)\\
\hline
Smearing & Gaussian smearing of electron energy and muon $p_T$ \\
\hline
Jet reconstruction &  {\tt PYCELL} cone algorithm in  {\tt PYTHIA}\\
\hline
Lepton-jet separation &  $\Delta R_{lj} \ge 0.4$ (for all jets)\\
\hline
Lepton-lepton separation & $\Delta R_{ll} \ge 0.2$\\
\hline
Lepton-photon separation & $\Delta R_{l\gamma} \ge 0.2$ for all ${p_T}_\gamma > 10~ \text{GeV}$\\ 
\hline
Hadronic activity & Hadronic activity for each lepton: \\ 
(To consider leptons with very less &  $\frac{\sum p_{T_{hadron}}}{p_{T_l}} \le 0.2$ ($\equiv$ radius of the cone around the \\
 hadronic activity around them.) & lepton) \\
\hline
Final $p_T$ cuts for leptons & ${p_T}_{l_1}>30$ GeV, ${p_T}_{l_2}>30$ GeV and ${p_T}_{l_3}>20$ GeV\\
\hline
Missing $p_T$ cut &  ${{\fmslash p}_T} > 30$ GeV\\
\hline
Z-veto \footnote{Invariant mass for the same flavored and opposite sign lepton pair, $m_{\ell_1
\ell_2}$, must be sufficiently away from $Z$ pole.} & $|m_{\ell_1\ell_2} - M_{Z}| \geq 6 \Gamma_{Z}$\\
\hline 
\hline
\multicolumn{2}{|c|}{\large VBF Cuts} \\
\hline
Central jet veto & Any additional jet with $p_T>20$ GeV, \\
&and $|\eta_0| < 2$, events are discarded\footnote{Pseudorapidity 
difference between the average of the two forward jets and the additional jet : $\eta_0 = \eta_3-(\eta_1+\eta2)/2$.}.\\
\hline
Pseudorapidity~\cite{Bozzi:2007ur} of charged leptons & $\eta_{j,min}<\eta_{\ell}<\eta_{j,max}$ \\
\hline
Cut applied to jets & ${p_T}_{j_1,j_2} > 20$ GeV\\
& $M_{j_1j_2} > 600$ GeV\\
&$\eta_{j_1}\cdot\eta_{j_2}<0$ and $|\eta_{j_1}-\eta_{j_2}|>4$\\\hline 
\end{tabular}
\caption{Selection criteria used in simulation. }
\label{table:selection}
\end{table}
%%%%%%%%%%%%%%%%%%%%%%%%%%%%%%%%%%%%%%%%%%%%%%%%

%%%%%%%%%%%%%%%%%%%%%%%%%%%%%%%%%%%%%%%%%%%%%%%%%%%%%%%%%%%%%%%%%%%%%%%%%%%%%%%%%%%%%%%%%%%%%%%%%%%%%%%%%%%%%%%%
\section{Simulation and Results}\label{sec:simulation_result}
%%%%%%%%%%%%%%%%%%%%%%%%%%%%%%%%%%%%%%%%%%%%%%%%%%%%%%%%%%%%%%%%%%%%%%%%%%%%%%%%%%%%%%%%%%%%%%%%%%%%%%%%%%%%%%%%
To analyze signals for heavy neutrino, we have implemented this model in {\tt FeynRules}~\cite{Christensen:2008py} 
to generate the Feynman rules compatible for {\tt MadGraph}. Parton level cross sections were generated using {\tt MadGraph5} and 
for showering and hadronization of the Les Houches Event~\cite{Alwall:2006yp} file, {\tt PYTHIA6}~\cite{Sjostrand:2006za} has been used. 

%%%%%%%%%%%%%%%%%%%%%%%%%%%%%%%%%%%%%%%%%%%%%%%%%%%%%%%%%%%%%%%%%%%%%%%%%%%%%%%%%%%%%%%%%%%%%%%%%%%%%%%%%%%%%%%%
%\subsection{Selection criteria}
%%%%%%%%%%%%%%%%%%%%%%%%%%%%%%%%%%%%%%%%%%%%%%%%%%%%%%%%%%%%%%%%%%%%%%%%%%%%%%%%%%%%%%%%%%%%%%%%%%%%%%%%%%%%%%%%
To enhance the signal over background, the selection criteria, tabulated in Table.~\ref{table:selection}, 
has been implemented. In top portion of this table, all selection parameters and efficiencies were listed. 
Cuts entitled with VBF cuts are applied only for VBF part of the analysis.
For detail see references~\cite{Bambhaniya:2013yca,Bambhaniya:2014kga}.

%%%%%%%%%%%%%%%%%%%%%%%%%%%%%%%%%%%%%%%%%%%%%%%%%%%%%%%%%%%%%%%%%%%%%%%%%%%%%%%%%%%%%%%%%%%%%%%%%%%%%%%%%%%%%%%%
%\subsection{Signal}
%%%%%%%%%%%%%%%%%%%%%%%%%%%%%%%%%%%%%%%%%%%%%%%%%%%%%%%%%%%%%%%%%%%%%%%%%%%%%%%%%%%%%%%%%%%%%%%%%%%%%%%%%%%%%%%%

Following from our earlier discussion on heavy neutrino production and decay, we are looking for tri-lepton production at the LHC,
$$p p \rightarrow \ell^{\pm} N  \rightarrow \ell^{\pm} (W^{\pm} \ell^{\mp}/Z \nu)  \rightarrow e^{\pm} e^{\pm} e^{\mp}
 / e^{\pm} \mu^{\pm} e^{\mp} / e^{\pm} \mu^{\pm} \mu^{\mp} /\mu^{\pm} \mu^{\pm} \mu^{\mp}+ \MET.$$

Cross section of final tri-lepton signal through $s$-channel heavy neutrino production at 14 TeV LHC for a benchmark point of
$M_R = 100$ GeV is listed in Table~\ref{table:signal_Schannel}. 
Here we have incorporated all event selection criteria except the VBF cuts.  
Total contribution from all the light leptons ($e,\,\mu$) as well as the differential contributions 
from the four flavor combinations are also presented. 
%%%%%%%%%%%%%%%%%%%%%%%%%%%%%%%%%%%%%%%%%%%%%%%%%%%%%%%%%%%%%%%%%%%%%%%%%%%%%%%%%%%%%%%%%%%%%%%%%%%%%%%%%%%%%%%%%%%%%%%%%%%%%
\begin{table}[t!]
\begin{tabular}{|c|c|}
\hline
  Total signal  &  Flavor allocated cross section ($fb$)\\
 cross section ($fb$) &~~~~eee ~~~~ $ee\mu$ ~~~~ $e\mu\mu$ ~~~ $\mu\mu\mu$\\
\hline
  2.732 & 0.318 ~~ 1.144 ~~ 1.030  ~~~ 0.2 \\
\hline
\end{tabular}
\caption{Final tri-lepton with $\MET$ signal cross section in $fb$ produced through 
$s$-channel heavy neutrino for the benchmark mass $M_R = 100$ GeV at the 14 
TeV LHC. All event selection cuts were applied (Table~\ref{table:selection}) except the VBF cuts as described in the text. 
We have also classified total tri-lepton signals into four different flavor combination of leptons and 
presented expected cross section in each category.}
\label{table:signal_Schannel}
\end{table}

%%%%%%%%%%%%%%%%%%%%%%%%%%%%%%%%%%%%%%%%%%%%%%%%%%%%%%%%%%%%%%%%%%%%%%%%%%%%%%%%%%%%%%%%%%%%%%%%%%%%%%%%%%%%%%%%
%\subsection{Background}
%%%%%%%%%%%%%%%%%%%%%%%%%%%%%%%%./%%%%%%%%%%%%%%%%%%%%%%%%%%%%%%%%%%%%%%%%%%%%%%%%%%%%%%%%%%%%%%%%%%%%%%%%%%%%%%%%

All the standard model channels those can mimic this tri-lepton signal with missing $E_T$ are considered for the estimation
of SM background. For such simulation events are generated using {\tt ALPGEN}~\cite{Mangano:2002ea} at
the parton level and then passed into {\tt PYTHIA} for hadronization and showering. 
We have used the same selection criteria as tabulated in Table~\ref{table:selection}. 
Inclusive cross section for $\ell^{\pm} \ell^{\pm} \ell^{\mp} \nu_{\ell}$ final state from the SM is $32.722\; fb$. 
Details of individual channel's contribution towards the SM background can be found
in ~\cite{Bambhaniya:2013wza,Bambhaniya:2013yca,Bambhaniya:2014kga}.

%%%%%%%%%%%%%%%%%%%%%%%%%%%%%%%%%%%%%%%%%%%%%%%%%%%%%%%%%%%%%%%%%%%%%%%%%%%%%%%%%%%%%%%%%%%%%%%%%%%%%%%%%%%%%%%%
\section{Discovery potential}\label{sec:discovery_pot}
%%%%%%%%%%%%%%%%%%%%%%%%%%%%%%%%%%%%%%%%%%%%%%%%%%%%%%%%%%%%%%%%%%%%%%%%%%%%%%%%%%%%%%%%%%%%%%%%%%%%%%%%%%%%%%%%
\begin{figure}[t!]
 \centering
  \includegraphics[angle=270,bb=50 50 554 770,scale=0.4,keepaspectratio=true]{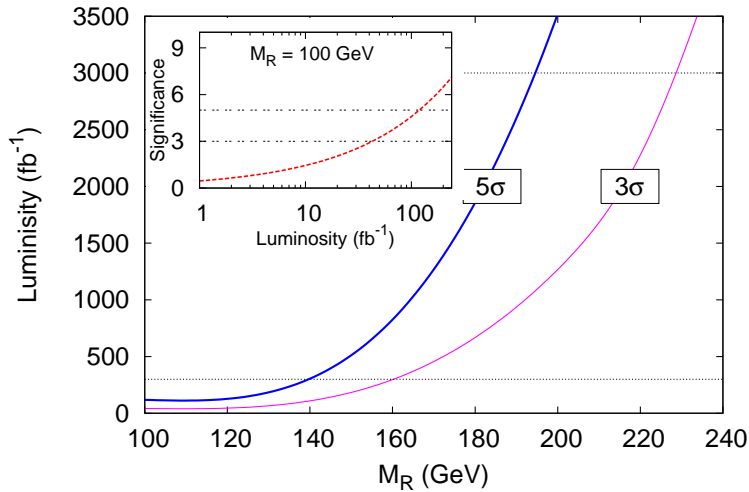}
 \caption{Contours of constant $3\sigma$ and $5\sigma$ significance at the 14 TeV LHC 
 in terms of heavy neutrino mass $M_R$ and integrated luminosity. 
With $300 fb^{-1}$ data tri-lepton signal can probe upto $M_R= 160(140)$ GeV with
$3\sigma(5\sigma)$ significance, whereas with $3000 fb^{-1}$ luminosity LHC can reach up to $230(190)$ GeV. 
Inset shows variation of significance for the $s$-channel tri-lepton production signal and backgrounds with 
heavy neutrino mass  $M_R=100$ GeV.}
 \label{fig:significance}
\end{figure}

With our understanding on signal strength of producing tri-leptons from heavy neutrino and possible sources of leading background, 
it is convenient to present our result  in terms of significance which we express as
${S}/{\sqrt{S+B}}$, where $S\,(B) = \mathcal{L} \,\sigma_{S\,(B)}$. $\mathcal{L}$ is the integrated luminosity of available data
from the experiment and $\sigma_{S(B)}$ is the final cross section of the signal (background) after all event selection 
cuts and  with model parameters of the model satisfying metastability and LFV bound. 
Fig.~\ref{fig:significance}  depicts $3\sigma$(magenta) and $5\sigma$(blue) constant  significance contours at the 
14 TeV LHC in terms of heavy neutrino mass and integrated luminosity. Horizontal black-dotted lines represent 
integrated luminosity of 300~$fb^{-1}$  and 3000~$fb^{-1}$. 
This model can be probed  through tri-lepton signals at the 14 TeV LHC upto $M_R= 160(140)$ GeV with
$3\sigma(5\sigma)$ significance with integrated luminosity of $300 fb^{-1}$. Whereas with higher luminosity of 
$3000 fb^{-1}$ it can be probed upto $\sim 230(190)$ GeV.
Inset of the figure demonstrates the expected significance of the $s$-channel tri-lepton production from heavy neutrino with 
mass $M_R=100$ GeV as a function of integrated luminosity. We note that $3 \,\sigma(5 \,\sigma)$ significance can be achieved 
with integrated luminosity $\sim 43(120) fb^{-1}$.

%%%%%%%%%%%%%%%%%%%%%%%%%%%%%%%%%%%%%%%%%%%%%%%%%%%%%%%%%%%%%%%%%%%%%%%%%%%%%%%%%%%%%%%%%%%%%%%%%%%%%%%%%%%%%%%%
\section{Conclusion}\label{sec:conclusion}
%%%%%%%%%%%%%%%%%%%%%%%%%%%%%%%%%%%%%%%%%%%%%%%%%%%%%%%%%%%%%%%%%%%%%%%%%%%%%%%%%%%%%%%%%%%%%%%%%%%%%%%%%%%%%%%%
In this work we have considered a TeV scale seesaw model that leads to quasi degenerate 
light neutrino mass spectrum. The model is fully reconstructible from oscillation parameters 
apart from an unknown factor parameterized by a constant $\omega$ for a common light and heavy neutrino mass scale, 
$m_\nu^d$ and $M_R$ respectively. 
We have demonstrated that the norm of Yukawa $\text{Tr}[Y_\nu^\dag Y_\nu]$ can choose arbitrary 
magnitude with different choices of $\omega$ and the common light neutrino mass scale $m_0$. 
Consequently we have obtained bounds on $\text{Tr}[Y_\nu^\dag Y_\nu]$ from both the 
consideration of the metastability of the electroweak vacuum as well as lepton flavor violation. 
Mass scale of QD light neutrinos are set at $m_0=0.07$ eV.
Extremely fine-tuned choices of unknown phases evade bound on $\text{Tr}[Y_\nu^\dag Y_\nu]$ from LFV. 
However bulk region of parameters allows us a stronger LFV bound than that of the metastability in the low $M_R$ regions.
Beyond that mass range, 
LFV bound becomes weaker than the metastability bound. The later remains slowly varying with $M_R$. 
However, contribution of the heavy neutrino towards the neutrino less double beta decay is 
insignificant in this model compared to the light neutrino contribution.

The constrained model parameters were then used to study the production and decay modes of 
the heavy neutrino at the LHC.  Due to suppressed same sign di-lepton signal in this model, 
we have studied tri-lepton associated with missing $E_T$ signal 
coming from the $s$-channel production of the heavy neutrino with realistic 
selection criteria as well as detailed simulation. However, the similar signal along with two 
forward tagged jets, coming through the production of heavy neutrino perceived in vector boson 
fusion comes with much smaller cross section at the present scenario.
With a benchmark point of heavy neutrino mass $M_R = 100$ GeV, we have presented the discovery 
potential of heavy neutrino, fitted to the model, with $3\,\sigma\,(5\,\sigma)$ significance for 
integrated luminosity $\sim 42(120)$ $fb^{-1}$ at the 14 TeV LHC. Moreover, this model can be probed for heavy 
neutrino mass upto $160 (230)$ GeV for low(high) luminosity options.

\vskip 1cm
%%%%%%%%%%%%%%%%%%%%%%%%%%%%%%%%%%%%%%%%%%%%%%%%%%%%%%%%%%%%%%%%%%%%%%%%%%%%%%%%%%%%%%%%%%%%%%%%%%%%
{\em Acknowledgements} \\
%%%%%%%%%%%%%%%%%%%%%%%%%%%%%%%%%%%%%%%%%%%%%%%%%%%%%%%%%%%%%%%%%%%%%%%%%%%%%%%%%%%%%%%%%%%%%%%%%%%%
We thank M. Ghosh, S. Goswami and S. Mohanty for useful discussions. P.K. also thanks KIAS group and KIAS workshop for hospitality.

\bibliographystyle{unsrt}
\bibliographystyle{apsrev4-1}
\bibliography{bibliography}

%Merlin.mbs v4.21 2009-07-09.
\begin{thebibliography}{10}%
\makeatletter
\providecommand \@ifxundefined [1]{%
 \ifx #1\undefined \expandafter \@firstoftwo
 \else \expandafter \@secondoftwo
\fi
}%
\providecommand \@ifnum [1]{%
 \ifnum #1\expandafter \@firstoftwo
 \else \expandafter \@secondoftwo
\fi
}%
\providecommand \enquote [1]{``#1''}%
\providecommand \bibnamefont  [1]{#1}%
\providecommand \bibfnamefont [1]{#1}%
\providecommand \citenamefont [1]{#1}%
\providecommand\href[0]{\@sanitize\@href}%
\providecommand\@href[1]{\endgroup\@@startlink{#1}\endgroup\@@href}%
\providecommand\@@href[1]{#1\@@endlink}%
\providecommand \@sanitize [0]{\begingroup\catcode`\&12\catcode`\#12\relax}%
\@ifxundefined \pdfoutput {\@firstoftwo}{%
 \@ifnum{\z@=\pdfoutput}{\@firstoftwo}{\@secondoftwo}%
}{%
 \providecommand\@@startlink[1]{\leavevmode\special{html:<a href="#1">}}%
 \providecommand\@@endlink[0]{\special{html:</a>}}%
}{%
 \providecommand\@@startlink[1]{%
  \leavevmode
  \pdfstartlink
   attr{/Border[0 0 1 ]/H/I/C[0 1 1]}%
   user{/Subtype/Link/A<</Type/Action/S/URI/URI(#1)>>}%
  \relax
 }%
 \providecommand\@@endlink[0]{\pdfendlink}%
}%
\providecommand \url  [0]{\begingroup\@sanitize \@url }%
\providecommand \@url [1]{\endgroup\@href {#1}{\urlprefix}}%
\providecommand \urlprefix [0]{URL }%
\providecommand \Eprint[0]{\href }%
\@ifxundefined \urlstyle {%
  \providecommand \doi [1]{doi:\discretionary{}{}{}#1}%
}{%
  \providecommand \doi [0]{doi:\discretionary{}{}{}\begingroup
  \urlstyle{rm}\Url }%
}%
\providecommand \doibase [0]{http://dx.doi.org/}%
\providecommand \Doi[1]{\href{\doibase#1}}%
\providecommand \bibAnnote [3]{%
  \BibitemShut{#1}%
  \begin{quotation}\noindent
    \textsc{Key:}\ #2\\\textsc{Annotation:}\ #3%
  \end{quotation}%
}%
\providecommand \bibAnnoteFile [2]{%
  \IfFileExists{#2}{\bibAnnote {#1} {#2} {\input{#2}}}{}%
}%
\providecommand \typeout [0]{\immediate \write \m@ne }%
\providecommand \selectlanguage [0]{\@gobble}%
\providecommand \bibinfo [0]{\@secondoftwo}%
\providecommand \bibfield [0]{\@secondoftwo}%
\providecommand \translation [1]{[#1]}%
\providecommand \BibitemOpen[0]{}%
\providecommand \bibitemStop [0]{}%
\providecommand \bibitemNoStop [0]{.\EOS\space}%
\providecommand \EOS [0]{\spacefactor3000\relax}%
\providecommand \BibitemShut [1]{\csname bibitem#1\endcsname}%
%</preamble>
\bibitem{:2012gu}%
  \BibitemOpen
  \bibfield{author}{%
  \bibinfo {author} {\bibfnamefont{S.}~\bibnamefont{Chatrchyan}} \emph{et~al.}
  (\bibinfo {collaboration} {CMS Collaboration}),\ }%
  \bibfield{journal}{%
  \Doi{10.1016/j.physletb.2012.08.021}{\bibinfo {journal} {Phys.Lett.}}\ }%
  \textbf{\bibinfo {volume} {B716}},\ \bibinfo {pages} {30} (\bibinfo {year}
  {2012}),\ \Eprint{http://arxiv.org/abs/1207.7235}{arXiv:1207.7235 [hep-ex]}%
  \bibAnnoteFile{NoStop}{:2012gu}%
%%CITATION = ARXIV:1207.7235;%%
\bibitem{:2012gk}%
  \BibitemOpen
  \bibfield{author}{%
  \bibinfo {author} {\bibfnamefont{G.}~\bibnamefont{Aad}} \emph{et~al.}
  (\bibinfo {collaboration} {ATLAS Collaboration}),\ }%
  \bibfield{journal}{%
  \Doi{10.1016/j.physletb.2012.08.020}{\bibinfo {journal} {Phys.Lett.}}\ }%
  \textbf{\bibinfo {volume} {B716}},\ \bibinfo {pages} {1} (\bibinfo {year}
  {2012}),\ \Eprint{http://arxiv.org/abs/1207.7214}{arXiv:1207.7214 [hep-ex]}%
  \bibAnnoteFile{NoStop}{:2012gk}%
%%CITATION = ARXIV:1207.7214;%%
\bibitem{Casas:1999cd}%
  \BibitemOpen
  \bibfield{author}{%
  \bibinfo {author} {\bibfnamefont{J.}~\bibnamefont{Casas}}, \bibinfo {author}
  {\bibfnamefont{V.}~\bibnamefont{Di~Clemente}}, \bibinfo {author}
  {\bibfnamefont{A.}~\bibnamefont{Ibarra}},\ and\ \bibinfo {author}
  {\bibfnamefont{M.}~\bibnamefont{Quiros}},\ }%
  \bibfield{journal}{%
  \Doi{10.1103/PhysRevD.62.053005}{\bibinfo {journal} {Phys.Rev.}}\ }%
  \textbf{\bibinfo {volume} {D62}},\ \bibinfo {pages} {053005} (\bibinfo {year}
  {2000}),\ \Eprint{http://arxiv.org/abs/hep-ph/9904295}{arXiv:hep-ph/9904295
  [hep-ph]}%
  \bibAnnoteFile{NoStop}{Casas:1999cd}%
%%CITATION = HEP-PH/9904295;%%
\bibitem{Gogoladze:2008ak}%
  \BibitemOpen
  \bibfield{author}{%
  \bibinfo {author} {\bibfnamefont{I.}~\bibnamefont{Gogoladze}}, \bibinfo
  {author} {\bibfnamefont{N.}~\bibnamefont{Okada}},\ and\ \bibinfo {author}
  {\bibfnamefont{Q.}~\bibnamefont{Shafi}},\ }%
  \bibfield{journal}{%
  \Doi{10.1016/j.physletb.2008.08.023}{\bibinfo {journal} {Phys.Lett.}}\ }%
  \textbf{\bibinfo {volume} {B668}},\ \bibinfo {pages} {121} (\bibinfo {year}
  {2008}),\ \Eprint{http://arxiv.org/abs/0805.2129}{arXiv:0805.2129 [hep-ph]}%
  \bibAnnoteFile{NoStop}{Gogoladze:2008ak}%
%%CITATION = ARXIV:0805.2129;%%
\bibitem{Shaposhnikov:2009pv}%
  \BibitemOpen
  \bibfield{author}{%
  \bibinfo {author} {\bibfnamefont{M.}~\bibnamefont{Shaposhnikov}}\ and\
  \bibinfo {author} {\bibfnamefont{C.}~\bibnamefont{Wetterich}},\ }%
  \bibfield{journal}{%
  \Doi{10.1016/j.physletb.2009.12.022}{\bibinfo {journal} {Phys.Lett.}}\ }%
  \textbf{\bibinfo {volume} {B683}},\ \bibinfo {pages} {196} (\bibinfo {year}
  {2010}),\ \Eprint{http://arxiv.org/abs/0912.0208}{arXiv:0912.0208 [hep-th]}%
  \bibAnnoteFile{NoStop}{Shaposhnikov:2009pv}%
%%CITATION = ARXIV:0912.0208;%%
\bibitem{Holthausen:2011aa}%
  \BibitemOpen
  \bibfield{author}{%
  \bibinfo {author} {\bibfnamefont{M.}~\bibnamefont{Holthausen}}, \bibinfo
  {author} {\bibfnamefont{K.~S.}\ \bibnamefont{Lim}},\ and\ \bibinfo {author}
  {\bibfnamefont{M.}~\bibnamefont{Lindner}},\ }%
  \bibfield{journal}{%
  \Doi{10.1007/JHEP02(2012)037}{\bibinfo {journal} {JHEP}}\ }%
  \textbf{\bibinfo {volume} {1202}},\ \bibinfo {pages} {037} (\bibinfo {year}
  {2012}),\ \Eprint{http://arxiv.org/abs/1112.2415}{arXiv:1112.2415 [hep-ph]}%
  \bibAnnoteFile{NoStop}{Holthausen:2011aa}%
%%CITATION = ARXIV:1112.2415;%%
\bibitem{Bezrukov:2012sa}%
  \BibitemOpen
  \bibfield{author}{%
  \bibinfo {author} {\bibfnamefont{F.}~\bibnamefont{Bezrukov}}, \bibinfo
  {author} {\bibfnamefont{M.~Y.}\ \bibnamefont{Kalmykov}}, \bibinfo {author}
  {\bibfnamefont{B.~A.}\ \bibnamefont{Kniehl}},\ and\ \bibinfo {author}
  {\bibfnamefont{M.}~\bibnamefont{Shaposhnikov}},\ }%
  \bibfield{journal}{%
  \Doi{10.1007/JHEP10(2012)140}{\bibinfo {journal} {JHEP}}\ }%
  \textbf{\bibinfo {volume} {1210}},\ \bibinfo {pages} {140} (\bibinfo {year}
  {2012}),\ \Eprint{http://arxiv.org/abs/1205.2893}{arXiv:1205.2893 [hep-ph]}%
  \bibAnnoteFile{NoStop}{Bezrukov:2012sa}%
%%CITATION = ARXIV:1205.2893;%%
\bibitem{Alekhin:2012py}%
  \BibitemOpen
  \bibfield{author}{%
  \bibinfo {author} {\bibfnamefont{S.}~\bibnamefont{Alekhin}}, \bibinfo
  {author} {\bibfnamefont{A.}~\bibnamefont{Djouadi}},\ and\ \bibinfo {author}
  {\bibfnamefont{S.}~\bibnamefont{Moch}},\ }%
  \bibfield{journal}{%
  \Doi{10.1016/j.physletb.2012.08.024}{\bibinfo {journal} {Phys.Lett.}}\ }%
  \textbf{\bibinfo {volume} {B716}},\ \bibinfo {pages} {214} (\bibinfo {year}
  {2012}),\ \Eprint{http://arxiv.org/abs/1207.0980}{arXiv:1207.0980 [hep-ph]}%
  \bibAnnoteFile{NoStop}{Alekhin:2012py}%
%%CITATION = ARXIV:1207.0980;%%
\bibitem{Degrassi:2012ry}%
  \BibitemOpen
  \bibfield{author}{%
  \bibinfo {author} {\bibfnamefont{G.}~\bibnamefont{Degrassi}}, \bibinfo
  {author} {\bibfnamefont{S.}~\bibnamefont{Di~Vita}}, \bibinfo {author}
  {\bibfnamefont{J.}~\bibnamefont{Elias-Miro}}, \bibinfo {author}
  {\bibfnamefont{J.~R.}\ \bibnamefont{Espinosa}}, \bibinfo {author}
  {\bibfnamefont{G.~F.}\ \bibnamefont{Giudice}}, \emph{et~al.},\ }%
  \bibfield{journal}{%
  \Doi{10.1007/JHEP08(2012)098}{\bibinfo {journal} {JHEP}}\ }%
  \textbf{\bibinfo {volume} {1208}},\ \bibinfo {pages} {098} (\bibinfo {year}
  {2012}),\ \Eprint{http://arxiv.org/abs/1205.6497}{arXiv:1205.6497 [hep-ph]}%
  \bibAnnoteFile{NoStop}{Degrassi:2012ry}%
%%CITATION = ARXIV:1205.6497;%%
\bibitem{Isidori:2001bm}%
  \BibitemOpen
  \bibfield{author}{%
  \bibinfo {author} {\bibfnamefont{G.}~\bibnamefont{Isidori}}, \bibinfo
  {author} {\bibfnamefont{G.}~\bibnamefont{Ridolfi}},\ and\ \bibinfo {author}
  {\bibfnamefont{A.}~\bibnamefont{Strumia}},\ }%
  \bibfield{journal}{%
  \Doi{10.1016/S0550-3213(01)00302-9}{\bibinfo {journal} {Nucl.Phys.}}\ }%
  \textbf{\bibinfo {volume} {B609}},\ \bibinfo {pages} {387} (\bibinfo {year}
  {2001}),\ \Eprint{http://arxiv.org/abs/hep-ph/0104016}{arXiv:hep-ph/0104016
  [hep-ph]}%
  \bibAnnoteFile{NoStop}{Isidori:2001bm}%
%%CITATION = HEP-PH/0104016;%%
\bibitem{Espinosa:2007qp}%
  \BibitemOpen
  \bibfield{author}{%
  \bibinfo {author} {\bibfnamefont{J.}~\bibnamefont{Espinosa}}, \bibinfo
  {author} {\bibfnamefont{G.}~\bibnamefont{Giudice}},\ and\ \bibinfo {author}
  {\bibfnamefont{A.}~\bibnamefont{Riotto}},\ }%
  \bibfield{journal}{%
  \Doi{10.1088/1475-7516/2008/05/002}{\bibinfo {journal} {JCAP}}\ }%
  \textbf{\bibinfo {volume} {0805}},\ \bibinfo {pages} {002} (\bibinfo {year}
  {2008}),\ \Eprint{http://arxiv.org/abs/0710.2484}{arXiv:0710.2484 [hep-ph]}%
  \bibAnnoteFile{NoStop}{Espinosa:2007qp}%
%%CITATION = ARXIV:0710.2484;%%
\bibitem{Ellis:2009tp}%
  \BibitemOpen
  \bibfield{author}{%
  \bibinfo {author} {\bibfnamefont{J.}~\bibnamefont{Ellis}}, \bibinfo {author}
  {\bibfnamefont{J.}~\bibnamefont{Espinosa}}, \bibinfo {author}
  {\bibfnamefont{G.}~\bibnamefont{Giudice}}, \bibinfo {author}
  {\bibfnamefont{A.}~\bibnamefont{Hoecker}},\ and\ \bibinfo {author}
  {\bibfnamefont{A.}~\bibnamefont{Riotto}},\ }%
  \bibfield{journal}{%
  \Doi{10.1016/j.physletb.2009.07.054}{\bibinfo {journal} {Phys.Lett.}}\ }%
  \textbf{\bibinfo {volume} {B679}},\ \bibinfo {pages} {369} (\bibinfo {year}
  {2009}),\ \Eprint{http://arxiv.org/abs/0906.0954}{arXiv:0906.0954 [hep-ph]}%
  \bibAnnoteFile{NoStop}{Ellis:2009tp}%
%%CITATION = ARXIV:0906.0954;%%
\bibitem{EliasMiro:2011aa}%
  \BibitemOpen
  \bibfield{author}{%
  \bibinfo {author} {\bibfnamefont{J.}~\bibnamefont{Elias-Miro}}, \bibinfo
  {author} {\bibfnamefont{J.~R.}\ \bibnamefont{Espinosa}}, \bibinfo {author}
  {\bibfnamefont{G.~F.}\ \bibnamefont{Giudice}}, \bibinfo {author}
  {\bibfnamefont{G.}~\bibnamefont{Isidori}}, \bibinfo {author}
  {\bibfnamefont{A.}~\bibnamefont{Riotto}}, \emph{et~al.},\ }%
  \bibfield{journal}{%
  \Doi{10.1016/j.physletb.2012.02.013}{\bibinfo {journal} {Phys.Lett.}}\ }%
  \textbf{\bibinfo {volume} {B709}},\ \bibinfo {pages} {222} (\bibinfo {year}
  {2012}),\ \Eprint{http://arxiv.org/abs/1112.3022}{arXiv:1112.3022 [hep-ph]}%
  \bibAnnoteFile{NoStop}{EliasMiro:2011aa}%
%%CITATION = ARXIV:1112.3022;%%
\bibitem{Rodejohann:2012px}%
  \BibitemOpen
  \bibfield{author}{%
  \bibinfo {author} {\bibfnamefont{W.}~\bibnamefont{Rodejohann}}\ and\ \bibinfo
  {author} {\bibfnamefont{H.}~\bibnamefont{Zhang}},\ }%
  \bibfield{journal}{%
  \Doi{10.1007/JHEP06(2012)022}{\bibinfo {journal} {JHEP}}\ }%
  \textbf{\bibinfo {volume} {1206}},\ \bibinfo {pages} {022} (\bibinfo {year}
  {2012}),\ \Eprint{http://arxiv.org/abs/1203.3825}{arXiv:1203.3825 [hep-ph]}%
  \bibAnnoteFile{NoStop}{Rodejohann:2012px}%
%%CITATION = ARXIV:1203.3825;%%
\bibitem{Chakrabortty:2012np}%
  \BibitemOpen
  \bibfield{author}{%
  \bibinfo {author} {\bibfnamefont{J.}~\bibnamefont{Chakrabortty}}, \bibinfo
  {author} {\bibfnamefont{M.}~\bibnamefont{Das}},\ and\ \bibinfo {author}
  {\bibfnamefont{S.}~\bibnamefont{Mohanty}},\ }%
  \bibfield{journal}{%
  \Doi{10.1142/S0217732313500326}{\bibinfo {journal} {Mod.Phys.Lett.}}\ }%
  \textbf{\bibinfo {volume} {A28}},\ \bibinfo {pages} {1350032} (\bibinfo
  {year} {2013}),\ \Eprint{http://arxiv.org/abs/1207.2027}{arXiv:1207.2027
  [hep-ph]}%
  \bibAnnoteFile{NoStop}{Chakrabortty:2012np}%
%%CITATION = ARXIV:1207.2027;%%
\bibitem{Chen:2012faa}%
  \BibitemOpen
  \bibfield{author}{%
  \bibinfo {author} {\bibfnamefont{C.-S.}\ \bibnamefont{Chen}}\ and\ \bibinfo
  {author} {\bibfnamefont{Y.}~\bibnamefont{Tang}},\ }%
  \bibfield{journal}{%
  \Doi{10.1007/JHEP04(2012)019}{\bibinfo {journal} {JHEP}}\ }%
  \textbf{\bibinfo {volume} {1204}},\ \bibinfo {pages} {019} (\bibinfo {year}
  {2012}),\ \Eprint{http://arxiv.org/abs/1202.5717}{arXiv:1202.5717 [hep-ph]}%
  \bibAnnoteFile{NoStop}{Chen:2012faa}%
%%CITATION = ARXIV:1202.5717;%%
\bibitem{Khan:2012zw}%
  \BibitemOpen
  \bibfield{author}{%
  \bibinfo {author} {\bibfnamefont{S.}~\bibnamefont{Khan}}, \bibinfo {author}
  {\bibfnamefont{S.}~\bibnamefont{Goswami}},\ and\ \bibinfo {author}
  {\bibfnamefont{S.}~\bibnamefont{Roy}},\ }%
  \bibfield{journal}{%
  \Doi{10.1103/PhysRevD.89.073021}{\bibinfo {journal} {Phys.Rev.}}\ }%
  \textbf{\bibinfo {volume} {D89}},\ \bibinfo {pages} {073021} (\bibinfo {year}
  {2014}),\ \Eprint{http://arxiv.org/abs/1212.3694}{arXiv:1212.3694 [hep-ph]}%
  \bibAnnoteFile{NoStop}{Khan:2012zw}%
%%CITATION = ARXIV:1212.3694;%%
\bibitem{Buttazzo:2013uya}%
  \BibitemOpen
  \bibfield{author}{%
  \bibinfo {author} {\bibfnamefont{D.}~\bibnamefont{Buttazzo}}, \bibinfo
  {author} {\bibfnamefont{G.}~\bibnamefont{Degrassi}}, \bibinfo {author}
  {\bibfnamefont{P.~P.}\ \bibnamefont{Giardino}}, \bibinfo {author}
  {\bibfnamefont{G.~F.}\ \bibnamefont{Giudice}}, \bibinfo {author}
  {\bibfnamefont{F.}~\bibnamefont{Sala}}, \emph{et~al.},\ }%
  \bibfield{journal}{%
  \Doi{10.1007/JHEP12(2013)089}{\bibinfo {journal} {JHEP}}\ }%
  \textbf{\bibinfo {volume} {1312}},\ \bibinfo {pages} {089} (\bibinfo {year}
  {2013}),\ \Eprint{http://arxiv.org/abs/1307.3536}{arXiv:1307.3536 [hep-ph]}%
  \bibAnnoteFile{NoStop}{Buttazzo:2013uya}%
%%CITATION = ARXIV:1307.3536;%%
\bibitem{Branchina:2013jra}%
  \BibitemOpen
  \bibfield{author}{%
  \bibinfo {author} {\bibfnamefont{V.}~\bibnamefont{Branchina}}\ and\ \bibinfo
  {author} {\bibfnamefont{E.}~\bibnamefont{Messina}},\ }%
  \bibfield{journal}{%
  \Doi{10.1103/PhysRevLett.111.241801}{\bibinfo {journal} {Phys.Rev.Lett.}}\ }%
  \textbf{\bibinfo {volume} {111}},\ \bibinfo {pages} {241801} (\bibinfo {year}
  {2013}),\ \Eprint{http://arxiv.org/abs/1307.5193}{arXiv:1307.5193 [hep-ph]}%
  \bibAnnoteFile{NoStop}{Branchina:2013jra}%
%%CITATION = ARXIV:1307.5193;%%
\bibitem{Branchina:2014usa}%
  \BibitemOpen
  \bibfield{author}{%
  \bibinfo {author} {\bibfnamefont{V.}~\bibnamefont{Branchina}}, \bibinfo
  {author} {\bibfnamefont{E.}~\bibnamefont{Messina}},\ and\ \bibinfo {author}
  {\bibfnamefont{A.}~\bibnamefont{Platania}},\ }%
  \bibfield{journal}{%
  \Doi{10.1007/JHEP09(2014)182}{\bibinfo {journal} {JHEP}}\ }%
  \textbf{\bibinfo {volume} {1409}},\ \bibinfo {pages} {182} (\bibinfo {year}
  {2014}),\ \Eprint{http://arxiv.org/abs/1407.4112}{arXiv:1407.4112 [hep-ph]}%
  \bibAnnoteFile{NoStop}{Branchina:2014usa}%
%%CITATION = ARXIV:1407.4112;%%
\bibitem{Branchina:2014rva}%
  \BibitemOpen
  \bibfield{author}{%
  \bibinfo {author} {\bibfnamefont{V.}~\bibnamefont{Branchina}}, \bibinfo
  {author} {\bibfnamefont{E.}~\bibnamefont{Messina}},\ and\ \bibinfo {author}
  {\bibfnamefont{M.}~\bibnamefont{Sher}},\ }%
  \bibfield{journal}{%
  \Doi{10.1103/PhysRevD.91.013003}{\bibinfo {journal} {Phys.Rev.}}\ }%
  \textbf{\bibinfo {volume} {D91}},\ \bibinfo {pages} {013003} (\bibinfo {year}
  {2015}),\ \Eprint{http://arxiv.org/abs/1408.5302}{arXiv:1408.5302 [hep-ph]}%
  \bibAnnoteFile{NoStop}{Branchina:2014rva}%
%%CITATION = ARXIV:1408.5302;%%
\bibitem{Petcov:2005jh}%
  \BibitemOpen
  \bibfield{author}{%
  \bibinfo {author} {\bibfnamefont{S.}~\bibnamefont{Petcov}}, \bibinfo {author}
  {\bibfnamefont{W.}~\bibnamefont{Rodejohann}}, \bibinfo {author}
  {\bibfnamefont{T.}~\bibnamefont{Shindou}},\ and\ \bibinfo {author}
  {\bibfnamefont{Y.}~\bibnamefont{Takanishi}},\ }%
  \bibfield{journal}{%
  \Doi{10.1016/j.nuclphysb.2006.01.034}{\bibinfo {journal} {Nucl.Phys.}}\ }%
  \textbf{\bibinfo {volume} {B739}},\ \bibinfo {pages} {208} (\bibinfo {year}
  {2006}),\ \Eprint{http://arxiv.org/abs/hep-ph/0510404}{arXiv:hep-ph/0510404
  [hep-ph]}%
  \bibAnnoteFile{NoStop}{Petcov:2005jh}%
%%CITATION = HEP-PH/0510404;%%
\bibitem{Dinh:2012bp}%
  \BibitemOpen
  \bibfield{author}{%
  \bibinfo {author} {\bibfnamefont{D.}~\bibnamefont{Dinh}}, \bibinfo {author}
  {\bibfnamefont{A.}~\bibnamefont{Ibarra}}, \bibinfo {author}
  {\bibfnamefont{E.}~\bibnamefont{Molinaro}},\ and\ \bibinfo {author}
  {\bibfnamefont{S.}~\bibnamefont{Petcov}},\ }%
  \bibfield{journal}{%
  \Doi{10.1007/JHEP08(2012)125}{\bibinfo {journal} {JHEP}}\ }%
  \textbf{\bibinfo {volume} {1208}},\ \bibinfo {pages} {125} (\bibinfo {year}
  {2012}),\ \Eprint{http://arxiv.org/abs/1205.4671}{arXiv:1205.4671 [hep-ph]}%
  \bibAnnoteFile{NoStop}{Dinh:2012bp}%
%%CITATION = ARXIV:1205.4671;%%
\bibitem{Abada:2014kba}%
  \BibitemOpen
  \bibfield{author}{%
  \bibinfo {author} {\bibfnamefont{A.}~\bibnamefont{Abada}}, \bibinfo {author}
  {\bibfnamefont{M.~E.}\ \bibnamefont{Krauss}}, \bibinfo {author}
  {\bibfnamefont{W.}~\bibnamefont{Porod}}, \bibinfo {author}
  {\bibfnamefont{F.}~\bibnamefont{Staub}}, \bibinfo {author}
  {\bibfnamefont{A.}~\bibnamefont{Vicente}}, \emph{et~al.},\ }%
  \bibfield{journal}{%
  \Doi{10.1007/JHEP11(2014)048}{\bibinfo {journal} {JHEP}}\ }%
  \textbf{\bibinfo {volume} {1411}},\ \bibinfo {pages} {048} (\bibinfo {year}
  {2014}),\ \Eprint{http://arxiv.org/abs/1408.0138}{arXiv:1408.0138 [hep-ph]}%
  \bibAnnoteFile{NoStop}{Abada:2014kba}%
%%CITATION = ARXIV:1408.0138;%%
\bibitem{Rodejohann:2012xd}%
  \BibitemOpen
  \bibfield{author}{%
  \bibinfo {author} {\bibfnamefont{W.}~\bibnamefont{Rodejohann}},\ }%
  \bibfield{journal}{%
  \Doi{10.1088/0954-3899/39/12/124008}{\bibinfo {journal} {J.Phys.}}\ }%
  \textbf{\bibinfo {volume} {G39}},\ \bibinfo {pages} {124008} (\bibinfo {year}
  {2012}),\ \Eprint{http://arxiv.org/abs/1206.2560}{arXiv:1206.2560 [hep-ph]}%
  \bibAnnoteFile{NoStop}{Rodejohann:2012xd}%
%%CITATION = ARXIV:1206.2560;%%
\bibitem{Bilenky:2014uka}%
  \BibitemOpen
  \bibfield{author}{%
  \bibinfo {author} {\bibfnamefont{S.}~\bibnamefont{Bilenky}}\ and\ \bibinfo
  {author} {\bibfnamefont{C.}~\bibnamefont{Giunti}},\ }%
  \bibfield{journal}{%
  \Doi{10.1142/S0217751X1530001X}{\bibinfo {journal} {Int.J.Mod.Phys.}}\ }%
  \textbf{\bibinfo {volume} {A30}},\ \bibinfo {pages} {1530001} (\bibinfo
  {year} {2015}),\ \Eprint{http://arxiv.org/abs/1411.4791}{arXiv:1411.4791
  [hep-ph]}%
  \bibAnnoteFile{NoStop}{Bilenky:2014uka}%
%%CITATION = ARXIV:1411.4791;%%
\bibitem{Huitu:1996su}%
  \BibitemOpen
  \bibfield{author}{%
  \bibinfo {author} {\bibfnamefont{K.}~\bibnamefont{Huitu}}, \bibinfo {author}
  {\bibfnamefont{J.}~\bibnamefont{Maalampi}}, \bibinfo {author}
  {\bibfnamefont{A.}~\bibnamefont{Pietila}},\ and\ \bibinfo {author}
  {\bibfnamefont{M.}~\bibnamefont{Raidal}},\ }%
  \bibfield{journal}{%
  \Doi{10.1016/S0550-3213(97)87466-4}{\bibinfo {journal} {Nucl.Phys.}}\ }%
  \textbf{\bibinfo {volume} {B487}},\ \bibinfo {pages} {27} (\bibinfo {year}
  {1997}),\ \Eprint{http://arxiv.org/abs/hep-ph/9606311}{arXiv:hep-ph/9606311
  [hep-ph]}%
  \bibAnnoteFile{NoStop}{Huitu:1996su}%
%%CITATION = HEP-PH/9606311;%%
\bibitem{Akeroyd:2005gt}%
  \BibitemOpen
  \bibfield{author}{%
  \bibinfo {author} {\bibfnamefont{A.}~\bibnamefont{Akeroyd}}\ and\ \bibinfo
  {author} {\bibfnamefont{M.}~\bibnamefont{Aoki}},\ }%
  \bibfield{journal}{%
  \Doi{10.1103/PhysRevD.72.035011}{\bibinfo {journal} {Phys.Rev.}}\ }%
  \textbf{\bibinfo {volume} {D72}},\ \bibinfo {pages} {035011} (\bibinfo {year}
  {2005}),\ \Eprint{http://arxiv.org/abs/hep-ph/0506176}{arXiv:hep-ph/0506176
  [hep-ph]}%
  \bibAnnoteFile{NoStop}{Akeroyd:2005gt}%
%%CITATION = HEP-PH/0506176;%%
\bibitem{Han:2006ip}%
  \BibitemOpen
  \bibfield{author}{%
  \bibinfo {author} {\bibfnamefont{T.}~\bibnamefont{Han}}\ and\ \bibinfo
  {author} {\bibfnamefont{B.}~\bibnamefont{Zhang}},\ }%
  \bibfield{journal}{%
  \Doi{10.1103/PhysRevLett.97.171804}{\bibinfo {journal} {Phys.Rev.Lett.}}\ }%
  \textbf{\bibinfo {volume} {97}},\ \bibinfo {pages} {171804} (\bibinfo {year}
  {2006}),\ \Eprint{http://arxiv.org/abs/hep-ph/0604064}{arXiv:hep-ph/0604064
  [hep-ph]}%
  \bibAnnoteFile{NoStop}{Han:2006ip}%
%%CITATION = HEP-PH/0604064;%%
\bibitem{delAguila:2007em}%
  \BibitemOpen
  \bibfield{author}{%
  \bibinfo {author} {\bibfnamefont{F.}~\bibnamefont{del Aguila}}, \bibinfo
  {author} {\bibfnamefont{J.}~\bibnamefont{Aguilar-Saavedra}},\ and\ \bibinfo
  {author} {\bibfnamefont{R.}~\bibnamefont{Pittau}},\ }%
  \bibfield{journal}{%
  \Doi{10.1088/1126-6708/2007/10/047}{\bibinfo {journal} {JHEP}}\ }%
  \textbf{\bibinfo {volume} {0710}},\ \bibinfo {pages} {047} (\bibinfo {year}
  {2007}),\ \Eprint{http://arxiv.org/abs/hep-ph/0703261}{arXiv:hep-ph/0703261
  [hep-ph]}%
  \bibAnnoteFile{NoStop}{delAguila:2007em}%
%%CITATION = HEP-PH/0703261;%%
\bibitem{Han:2007bk}%
  \BibitemOpen
  \bibfield{author}{%
  \bibinfo {author} {\bibfnamefont{T.}~\bibnamefont{Han}}, \bibinfo {author}
  {\bibfnamefont{B.}~\bibnamefont{Mukhopadhyaya}}, \bibinfo {author}
  {\bibfnamefont{Z.}~\bibnamefont{Si}},\ and\ \bibinfo {author}
  {\bibfnamefont{K.}~\bibnamefont{Wang}},\ }%
  \bibfield{journal}{%
  \Doi{10.1103/PhysRevD.76.075013}{\bibinfo {journal} {Phys.Rev.}}\ }%
  \textbf{\bibinfo {volume} {D76}},\ \bibinfo {pages} {075013} (\bibinfo {year}
  {2007}),\ \Eprint{http://arxiv.org/abs/0706.0441}{arXiv:0706.0441 [hep-ph]}%
  \bibAnnoteFile{NoStop}{Han:2007bk}%
%%CITATION = ARXIV:0706.0441;%%
\bibitem{Akeroyd:2007zv}%
  \BibitemOpen
  \bibfield{author}{%
  \bibinfo {author} {\bibfnamefont{A.}~\bibnamefont{Akeroyd}}, \bibinfo
  {author} {\bibfnamefont{M.}~\bibnamefont{Aoki}},\ and\ \bibinfo {author}
  {\bibfnamefont{H.}~\bibnamefont{Sugiyama}},\ }%
  \bibfield{journal}{%
  \Doi{10.1103/PhysRevD.77.075010}{\bibinfo {journal} {Phys.Rev.}}\ }%
  \textbf{\bibinfo {volume} {D77}},\ \bibinfo {pages} {075010} (\bibinfo {year}
  {2008}),\ \Eprint{http://arxiv.org/abs/0712.4019}{arXiv:0712.4019 [hep-ph]}%
  \bibAnnoteFile{NoStop}{Akeroyd:2007zv}%
%%CITATION = ARXIV:0712.4019;%%
\bibitem{Bray:2007ru}%
  \BibitemOpen
  \bibfield{author}{%
  \bibinfo {author} {\bibfnamefont{S.}~\bibnamefont{Bray}}, \bibinfo {author}
  {\bibfnamefont{J.~S.}\ \bibnamefont{Lee}},\ and\ \bibinfo {author}
  {\bibfnamefont{A.}~\bibnamefont{Pilaftsis}},\ }%
  \bibfield{journal}{%
  \Doi{10.1016/j.nuclphysb.2007.07.002}{\bibinfo {journal} {Nucl.Phys.}}\ }%
  \textbf{\bibinfo {volume} {B786}},\ \bibinfo {pages} {95} (\bibinfo {year}
  {2007}),\ \Eprint{http://arxiv.org/abs/hep-ph/0702294}{arXiv:hep-ph/0702294
  [HEP-PH]}%
  \bibAnnoteFile{NoStop}{Bray:2007ru}%
%%CITATION = HEP-PH/0702294;%%
\bibitem{Perez:2008ha}%
  \BibitemOpen
  \bibfield{author}{%
  \bibinfo {author} {\bibfnamefont{P.}~\bibnamefont{Fileviez~Perez}}, \bibinfo
  {author} {\bibfnamefont{T.}~\bibnamefont{Han}}, \bibinfo {author}
  {\bibfnamefont{G.-y.}\ \bibnamefont{Huang}}, \bibinfo {author}
  {\bibfnamefont{T.}~\bibnamefont{Li}},\ and\ \bibinfo {author}
  {\bibfnamefont{K.}~\bibnamefont{Wang}},\ }%
  \bibfield{journal}{%
  \Doi{10.1103/PhysRevD.78.015018}{\bibinfo {journal} {Phys.Rev.}}\ }%
  \textbf{\bibinfo {volume} {D78}},\ \bibinfo {pages} {015018} (\bibinfo {year}
  {2008}),\ \Eprint{http://arxiv.org/abs/0805.3536}{arXiv:0805.3536 [hep-ph]}%
  \bibAnnoteFile{NoStop}{Perez:2008ha}%
%%CITATION = ARXIV:0805.3536;%%
\bibitem{delAguila:2008cj}%
  \BibitemOpen
  \bibfield{author}{%
  \bibinfo {author} {\bibfnamefont{F.}~\bibnamefont{del Aguila}}\ and\ \bibinfo
  {author} {\bibfnamefont{J.}~\bibnamefont{Aguilar-Saavedra}},\ }%
  \bibfield{journal}{%
  \Doi{10.1016/j.nuclphysb.2008.12.029}{\bibinfo {journal} {Nucl.Phys.}}\ }%
  \textbf{\bibinfo {volume} {B813}},\ \bibinfo {pages} {22} (\bibinfo {year}
  {2009}),\ \Eprint{http://arxiv.org/abs/0808.2468}{arXiv:0808.2468 [hep-ph]}%
  \bibAnnoteFile{NoStop}{delAguila:2008cj}%
%%CITATION = ARXIV:0808.2468;%%
\bibitem{Franceschini:2008pz}%
  \BibitemOpen
  \bibfield{author}{%
  \bibinfo {author} {\bibfnamefont{R.}~\bibnamefont{Franceschini}}, \bibinfo
  {author} {\bibfnamefont{T.}~\bibnamefont{Hambye}},\ and\ \bibinfo {author}
  {\bibfnamefont{A.}~\bibnamefont{Strumia}},\ }%
  \bibfield{journal}{%
  \Doi{10.1103/PhysRevD.78.033002}{\bibinfo {journal} {Phys.Rev.}}\ }%
  \textbf{\bibinfo {volume} {D78}},\ \bibinfo {pages} {033002} (\bibinfo {year}
  {2008}),\ \Eprint{http://arxiv.org/abs/0805.1613}{arXiv:0805.1613 [hep-ph]}%
  \bibAnnoteFile{NoStop}{Franceschini:2008pz}%
%%CITATION = ARXIV:0805.1613;%%
\bibitem{Xing:2009mm}%
  \BibitemOpen
  \bibfield{author}{%
  \bibinfo {author} {\bibfnamefont{Z.-Z.}\ \bibnamefont{Xing}},\ }%
  \bibfield{journal}{%
  \Doi{10.1142/S0217751X09046886}{\bibinfo {journal} {Int.J.Mod.Phys.}}\ }%
  \textbf{\bibinfo {volume} {A24}},\ \bibinfo {pages} {3286} (\bibinfo {year}
  {2009}),\ \Eprint{http://arxiv.org/abs/0901.0209}{arXiv:0901.0209 [hep-ph]}%
  \bibAnnoteFile{NoStop}{Xing:2009mm}%
%%CITATION = ARXIV:0901.0209;%%
\bibitem{Atre:2009rg}%
  \BibitemOpen
  \bibfield{author}{%
  \bibinfo {author} {\bibfnamefont{A.}~\bibnamefont{Atre}}, \bibinfo {author}
  {\bibfnamefont{T.}~\bibnamefont{Han}}, \bibinfo {author}
  {\bibfnamefont{S.}~\bibnamefont{Pascoli}},\ and\ \bibinfo {author}
  {\bibfnamefont{B.}~\bibnamefont{Zhang}},\ }%
  \bibfield{journal}{%
  \Doi{10.1088/1126-6708/2009/05/030}{\bibinfo {journal} {JHEP}}\ }%
  \textbf{\bibinfo {volume} {0905}},\ \bibinfo {pages} {030} (\bibinfo {year}
  {2009}),\ \Eprint{http://arxiv.org/abs/0901.3589}{arXiv:0901.3589 [hep-ph]}%
  \bibAnnoteFile{NoStop}{Atre:2009rg}%
%%CITATION = ARXIV:0901.3589;%%
\bibitem{Melfo:2011nx}%
  \BibitemOpen
  \bibfield{author}{%
  \bibinfo {author} {\bibfnamefont{A.}~\bibnamefont{Melfo}}, \bibinfo {author}
  {\bibfnamefont{M.}~\bibnamefont{Nemevsek}}, \bibinfo {author}
  {\bibfnamefont{F.}~\bibnamefont{Nesti}}, \bibinfo {author}
  {\bibfnamefont{G.}~\bibnamefont{Senjanovic}},\ and\ \bibinfo {author}
  {\bibfnamefont{Y.}~\bibnamefont{Zhang}},\ }%
  \bibfield{journal}{%
  \Doi{10.1103/PhysRevD.85.055018}{\bibinfo {journal} {Phys.Rev.}}\ }%
  \textbf{\bibinfo {volume} {D85}},\ \bibinfo {pages} {055018} (\bibinfo {year}
  {2012}),\ \Eprint{http://arxiv.org/abs/1108.4416}{arXiv:1108.4416 [hep-ph]}%
  \bibAnnoteFile{NoStop}{Melfo:2011nx}%
%%CITATION = ARXIV:1108.4416;%%
\bibitem{Chen:2011de}%
  \BibitemOpen
  \bibfield{author}{%
  \bibinfo {author} {\bibfnamefont{M.-C.}\ \bibnamefont{Chen}}\ and\ \bibinfo
  {author} {\bibfnamefont{J.}~\bibnamefont{Huang}},\ }%
  \bibfield{journal}{%
  \Doi{10.1142/S0217732311035985}{\bibinfo {journal} {Mod.Phys.Lett.}}\ }%
  \textbf{\bibinfo {volume} {A26}},\ \bibinfo {pages} {1147} (\bibinfo {year}
  {2011}),\ \Eprint{http://arxiv.org/abs/1105.3188}{arXiv:1105.3188 [hep-ph]}%
  \bibAnnoteFile{NoStop}{Chen:2011de}%
%%CITATION = ARXIV:1105.3188;%%
\bibitem{Eboli:2011ia}%
  \BibitemOpen
  \bibfield{author}{%
  \bibinfo {author} {\bibfnamefont{O.}~\bibnamefont{Eboli}}, \bibinfo {author}
  {\bibfnamefont{J.}~\bibnamefont{Gonzalez-Fraile}},\ and\ \bibinfo {author}
  {\bibfnamefont{M.}~\bibnamefont{Gonzalez-Garcia}},\ }%
  \bibfield{journal}{%
  \Doi{10.1007/JHEP12(2011)009}{\bibinfo {journal} {JHEP}}\ }%
  \textbf{\bibinfo {volume} {1112}},\ \bibinfo {pages} {009} (\bibinfo {year}
  {2011}),\ \Eprint{http://arxiv.org/abs/1108.0661}{arXiv:1108.0661 [hep-ph]}%
  \bibAnnoteFile{NoStop}{Eboli:2011ia}%
%%CITATION = ARXIV:1108.0661;%%
\bibitem{Vanini:2012mla}%
  \BibitemOpen
  \bibfield{author}{%
  \bibinfo {author} {\bibfnamefont{S.}~\bibnamefont{Vanini}}}%
   (\bibinfo {year} {2012}),\ \url{https://inspirehep.net/record/1231294}%
  \bibAnnoteFile{NoStop}{Vanini:2012mla}%
%%CITATION = CERN-THESIS-2012-175 ETC.;%%
\bibitem{Das:2012ze}%
  \BibitemOpen
  \bibfield{author}{%
  \bibinfo {author} {\bibfnamefont{A.}~\bibnamefont{Das}}\ and\ \bibinfo
  {author} {\bibfnamefont{N.}~\bibnamefont{Okada}},\ }%
  \bibfield{journal}{%
  \Doi{10.1103/PhysRevD.88.113001}{\bibinfo {journal} {Phys.Rev.}}\ }%
  \textbf{\bibinfo {volume} {D88}},\ \bibinfo {pages} {113001} (\bibinfo {year}
  {2013}),\ \Eprint{http://arxiv.org/abs/1207.3734}{arXiv:1207.3734 [hep-ph]}%
  \bibAnnoteFile{NoStop}{Das:2012ze}%
%%CITATION = ARXIV:1207.3734;%%
\bibitem{Bambhaniya:2013yca}%
  \BibitemOpen
  \bibfield{author}{%
  \bibinfo {author} {\bibfnamefont{G.}~\bibnamefont{Bambhaniya}}, \bibinfo
  {author} {\bibfnamefont{J.}~\bibnamefont{Chakrabortty}}, \bibinfo {author}
  {\bibfnamefont{S.}~\bibnamefont{Goswami}},\ and\ \bibinfo {author}
  {\bibfnamefont{P.}~\bibnamefont{Konar}},\ }%
  \bibfield{journal}{%
  \Doi{10.1103/PhysRevD.88.075006}{\bibinfo {journal} {Phys.Rev.}}\ }%
  \textbf{\bibinfo {volume} {D88}},\ \bibinfo {pages} {075006} (\bibinfo {year}
  {2013}),\ \Eprint{http://arxiv.org/abs/1305.2795}{arXiv:1305.2795 [hep-ph]}%
  \bibAnnoteFile{NoStop}{Bambhaniya:2013yca}%
%%CITATION = ARXIV:1305.2795;%%
\bibitem{Dev:2013wba}%
  \BibitemOpen
  \bibfield{author}{%
  \bibinfo {author} {\bibfnamefont{P.~S.~B.}\ \bibnamefont{Dev}}, \bibinfo
  {author} {\bibfnamefont{A.}~\bibnamefont{Pilaftsis}},\ and\ \bibinfo {author}
  {\bibfnamefont{U.-k.}\ \bibnamefont{Yang}},\ }%
  \bibfield{journal}{%
  \Doi{10.1103/PhysRevLett.112.081801}{\bibinfo {journal} {Phys.Rev.Lett.}}\ }%
  \textbf{\bibinfo {volume} {112}},\ \bibinfo {pages} {081801} (\bibinfo {year}
  {2014}),\ \Eprint{http://arxiv.org/abs/1308.2209}{arXiv:1308.2209 [hep-ph]}%
  \bibAnnoteFile{NoStop}{Dev:2013wba}%
%%CITATION = ARXIV:1308.2209;%%
\bibitem{Aguilar-Saavedra:2013twa}%
  \BibitemOpen
  \bibfield{author}{%
  \bibinfo {author} {\bibfnamefont{J.}~\bibnamefont{Aguilar-Saavedra}},
  \bibinfo {author} {\bibfnamefont{P.}~\bibnamefont{Boavida}},\ and\ \bibinfo
  {author} {\bibfnamefont{F.}~\bibnamefont{Joaquim}},\ }%
  \bibfield{journal}{%
  \Doi{10.1103/PhysRevD.88.113008}{\bibinfo {journal} {Phys.Rev.}}\ }%
  \textbf{\bibinfo {volume} {D88}},\ \bibinfo {pages} {113008} (\bibinfo {year}
  {2013}),\ \Eprint{http://arxiv.org/abs/1308.3226}{arXiv:1308.3226 [hep-ph]}%
  \bibAnnoteFile{NoStop}{Aguilar-Saavedra:2013twa}%
%%CITATION = ARXIV:1308.3226;%%
\bibitem{Das:2014jxa}%
  \BibitemOpen
  \bibfield{author}{%
  \bibinfo {author} {\bibfnamefont{A.}~\bibnamefont{Das}}, \bibinfo {author}
  {\bibfnamefont{P.}~\bibnamefont{Bhupal~Dev}},\ and\ \bibinfo {author}
  {\bibfnamefont{N.}~\bibnamefont{Okada}},\ }%
  \bibfield{journal}{%
  \Doi{10.1016/j.physletb.2014.06.058}{\bibinfo {journal} {Phys.Lett.}}\ }%
  \textbf{\bibinfo {volume} {B735}},\ \bibinfo {pages} {364} (\bibinfo {year}
  {2014}),\ \Eprint{http://arxiv.org/abs/1405.0177}{arXiv:1405.0177 [hep-ph]}%
  \bibAnnoteFile{NoStop}{Das:2014jxa}%
%%CITATION = ARXIV:1405.0177;%%
\bibitem{Bambhaniya:2014kga}%
  \BibitemOpen
  \bibfield{author}{%
  \bibinfo {author} {\bibfnamefont{G.}~\bibnamefont{Bambhaniya}}, \bibinfo
  {author} {\bibfnamefont{S.}~\bibnamefont{Goswami}}, \bibinfo {author}
  {\bibfnamefont{S.}~\bibnamefont{Khan}}, \bibinfo {author}
  {\bibfnamefont{P.}~\bibnamefont{Konar}},\ and\ \bibinfo {author}
  {\bibfnamefont{T.}~\bibnamefont{Mondal}},\ }%
  \bibfield{journal}{%
  \Doi{10.1103/PhysRevD.91.075007}{\bibinfo {journal} {Phys.Rev.}}\ }%
  \textbf{\bibinfo {volume} {D91}},\ \bibinfo {pages} {075007} (\bibinfo {year}
  {2015}),\ \Eprint{http://arxiv.org/abs/1410.5687}{arXiv:1410.5687 [hep-ph]}%
  \bibAnnoteFile{NoStop}{Bambhaniya:2014kga}%
%%CITATION = ARXIV:1410.5687;%%
\bibitem{Minkowski:1977sc}%
  \BibitemOpen
  \bibfield{author}{%
  \bibinfo {author} {\bibfnamefont{P.}~\bibnamefont{Minkowski}},\ }%
  \bibfield{journal}{%
  \Doi{10.1016/0370-2693(77)90435-X}{\bibinfo {journal} {Phys.Lett.}}\ }%
  \textbf{\bibinfo {volume} {B67}},\ \bibinfo {pages} {421} (\bibinfo {year}
  {1977})%
  \bibAnnoteFile{NoStop}{Minkowski:1977sc}%
%%CITATION = PHLTA,B67,421;%%
\bibitem{Yanagida:1979as}%
  \BibitemOpen
  \bibfield{author}{%
  \bibinfo {author} {\bibfnamefont{T.}~\bibnamefont{Yanagida}},\ }%
  \bibfield{journal}{%
  \bibinfo {journal} {Conf.Proc.}\ }%
  \textbf{\bibinfo {volume} {C7902131}},\ \bibinfo {pages} {95} (\bibinfo
  {year} {1979})%
  \bibAnnoteFile{NoStop}{Yanagida:1979as}%
%%CITATION = CONFP,C7902131,95;%%
\bibitem{GellMann:1980vs}%
  \BibitemOpen
  \bibfield{author}{%
  \bibinfo {author} {\bibfnamefont{M.}~\bibnamefont{Gell-Mann}}, \bibinfo
  {author} {\bibfnamefont{P.}~\bibnamefont{Ramond}},\ and\ \bibinfo {author}
  {\bibfnamefont{R.}~\bibnamefont{Slansky}},\ }%
  \bibfield{journal}{%
  \bibinfo {journal} {Conf.Proc.}\ }%
  \textbf{\bibinfo {volume} {C790927}},\ \bibinfo {pages} {315} (\bibinfo
  {year} {1979}),\ \Eprint{http://arxiv.org/abs/1306.4669}{arXiv:1306.4669
  [hep-th]}%
  \bibAnnoteFile{NoStop}{GellMann:1980vs}%
%%CITATION = ARXIV:1306.4669;%%
\bibitem{Glashow:1979nm}%
  \BibitemOpen
  \bibfield{author}{%
  \bibinfo {author} {\bibfnamefont{S.}~\bibnamefont{Glashow}},\ }%
  \bibfield{journal}{%
  \bibinfo {journal} {NATO Adv.Study Inst.Ser.B Phys.}\ }%
  \textbf{\bibinfo {volume} {59}},\ \bibinfo {pages} {687} (\bibinfo {year}
  {1980})%
  \bibAnnoteFile{NoStop}{Glashow:1979nm}%
%%CITATION = NASBD,59,687;%%
\bibitem{Mohapatra:1979ia}%
  \BibitemOpen
  \bibfield{author}{%
  \bibinfo {author} {\bibfnamefont{R.~N.}\ \bibnamefont{Mohapatra}}\ and\
  \bibinfo {author} {\bibfnamefont{G.}~\bibnamefont{Senjanovic}},\ }%
  \bibfield{journal}{%
  \Doi{10.1103/PhysRevLett.44.912}{\bibinfo {journal} {Phys.Rev.Lett.}}\ }%
  \textbf{\bibinfo {volume} {44}},\ \bibinfo {pages} {912} (\bibinfo {year}
  {1980})%
  \bibAnnoteFile{NoStop}{Mohapatra:1979ia}%
%%CITATION = PRLTA,44,912;%%
\bibitem{Schechter:1980gr}%
  \BibitemOpen
  \bibfield{author}{%
  \bibinfo {author} {\bibfnamefont{J.}~\bibnamefont{Schechter}}\ and\ \bibinfo
  {author} {\bibfnamefont{J.}~\bibnamefont{Valle}},\ }%
  \bibfield{journal}{%
  \Doi{10.1103/PhysRevD.22.2227}{\bibinfo {journal} {Phys.Rev.}}\ }%
  \textbf{\bibinfo {volume} {D22}},\ \bibinfo {pages} {2227} (\bibinfo {year}
  {1980})%
  \bibAnnoteFile{NoStop}{Schechter:1980gr}%
%%CITATION = PHRVA,D22,2227;%%
\bibitem{Ade:2013zuv}%
  \BibitemOpen
  \bibfield{author}{%
  \bibinfo {author} {\bibfnamefont{P.}~\bibnamefont{Ade}} \emph{et~al.}
  (\bibinfo {collaboration} {Planck}),\ }%
  \bibfield{journal}{%
  \Doi{10.1051/0004-6361/201321591}{\bibinfo {journal} {Astron.Astrophys.}}\ }%
  \textbf{\bibinfo {volume} {571}},\ \bibinfo {pages} {A16} (\bibinfo {year}
  {2014}),\ \Eprint{http://arxiv.org/abs/1303.5076}{arXiv:1303.5076
  [astro-ph.CO]}%
  \bibAnnoteFile{NoStop}{Ade:2013zuv}%
%%CITATION = ARXIV:1303.5076;%%
\bibitem{Rainwater:1999gg}%
  \BibitemOpen
  \bibfield{author}{%
  \bibinfo {author} {\bibfnamefont{D.~L.}\ \bibnamefont{Rainwater}}}%
   (\bibinfo {year} {1999}),\
  \Eprint{http://arxiv.org/abs/hep-ph/9908378}{arXiv:hep-ph/9908378 [hep-ph]}%
  \bibAnnoteFile{NoStop}{Rainwater:1999gg}%
%%CITATION = HEP-PH/9908378;%%
\bibitem{Datta:2001cy}%
  \BibitemOpen
  \bibfield{author}{%
  \bibinfo {author} {\bibfnamefont{A.}~\bibnamefont{Datta}}, \bibinfo {author}
  {\bibfnamefont{P.}~\bibnamefont{Konar}},\ and\ \bibinfo {author}
  {\bibfnamefont{B.}~\bibnamefont{Mukhopadhyaya}},\ }%
  \bibfield{journal}{%
  \Doi{10.1103/PhysRevD.65.055008}{\bibinfo {journal} {Phys.Rev.}}\ }%
  \textbf{\bibinfo {volume} {D65}},\ \bibinfo {pages} {055008} (\bibinfo {year}
  {2002}),\ \Eprint{http://arxiv.org/abs/hep-ph/0109071}{arXiv:hep-ph/0109071
  [hep-ph]}%
  \bibAnnoteFile{NoStop}{Datta:2001cy}%
%%CITATION = HEP-PH/0109071;%%
\bibitem{Datta:2001hv}%
  \BibitemOpen
  \bibfield{author}{%
  \bibinfo {author} {\bibfnamefont{A.}~\bibnamefont{Datta}}, \bibinfo {author}
  {\bibfnamefont{P.}~\bibnamefont{Konar}},\ and\ \bibinfo {author}
  {\bibfnamefont{B.}~\bibnamefont{Mukhopadhyaya}},\ }%
  \bibfield{journal}{%
  \Doi{10.1103/PhysRevLett.88.181802}{\bibinfo {journal} {Phys.Rev.Lett.}}\ }%
  \textbf{\bibinfo {volume} {88}},\ \bibinfo {pages} {181802} (\bibinfo {year}
  {2002}),\ \Eprint{http://arxiv.org/abs/hep-ph/0111012}{arXiv:hep-ph/0111012
  [hep-ph]}%
  \bibAnnoteFile{NoStop}{Datta:2001hv}%
%%CITATION = HEP-PH/0111012;%%
\bibitem{Choudhury:2003hq}%
  \BibitemOpen
  \bibfield{author}{%
  \bibinfo {author} {\bibfnamefont{D.}~\bibnamefont{Choudhury}}, \bibinfo
  {author} {\bibfnamefont{A.}~\bibnamefont{Datta}}, \bibinfo {author}
  {\bibfnamefont{K.}~\bibnamefont{Huitu}}, \bibinfo {author}
  {\bibfnamefont{P.}~\bibnamefont{Konar}}, \bibinfo {author}
  {\bibfnamefont{S.}~\bibnamefont{Moretti}}, \emph{et~al.},\ }%
  \bibfield{journal}{%
  \Doi{10.1103/PhysRevD.68.075007}{\bibinfo {journal} {Phys.Rev.}}\ }%
  \textbf{\bibinfo {volume} {D68}},\ \bibinfo {pages} {075007} (\bibinfo {year}
  {2003}),\ \Eprint{http://arxiv.org/abs/hep-ph/0304192}{arXiv:hep-ph/0304192
  [hep-ph]}%
  \bibAnnoteFile{NoStop}{Choudhury:2003hq}%
%%CITATION = HEP-PH/0304192;%%
\bibitem{Cho:2006sx}%
  \BibitemOpen
  \bibfield{author}{%
  \bibinfo {author} {\bibfnamefont{G.-C.}\ \bibnamefont{Cho}}, \bibinfo
  {author} {\bibfnamefont{K.}~\bibnamefont{Hagiwara}}, \bibinfo {author}
  {\bibfnamefont{J.}~\bibnamefont{Kanzaki}}, \bibinfo {author}
  {\bibfnamefont{T.}~\bibnamefont{Plehn}}, \bibinfo {author}
  {\bibfnamefont{D.}~\bibnamefont{Rainwater}}, \emph{et~al.},\ }%
  \bibfield{journal}{%
  \Doi{10.1103/PhysRevD.73.054002}{\bibinfo {journal} {Phys.Rev.}}\ }%
  \textbf{\bibinfo {volume} {D73}},\ \bibinfo {pages} {054002} (\bibinfo {year}
  {2006}),\ \Eprint{http://arxiv.org/abs/hep-ph/0601063}{arXiv:hep-ph/0601063
  [hep-ph]}%
  \bibAnnoteFile{NoStop}{Cho:2006sx}%
%%CITATION = HEP-PH/0601063;%%
\bibitem{Casas:2001sr}%
  \BibitemOpen
  \bibfield{author}{%
  \bibinfo {author} {\bibfnamefont{J.}~\bibnamefont{Casas}}\ and\ \bibinfo
  {author} {\bibfnamefont{A.}~\bibnamefont{Ibarra}},\ }%
  \bibfield{journal}{%
  \Doi{10.1016/S0550-3213(01)00475-8}{\bibinfo {journal} {Nucl.Phys.}}\ }%
  \textbf{\bibinfo {volume} {B618}},\ \bibinfo {pages} {171} (\bibinfo {year}
  {2001}),\ \Eprint{http://arxiv.org/abs/hep-ph/0103065}{arXiv:hep-ph/0103065
  [hep-ph]}%
  \bibAnnoteFile{NoStop}{Casas:2001sr}%
%%CITATION = HEP-PH/0103065;%%
\bibitem{Ibarra:2003up}%
  \BibitemOpen
  \bibfield{author}{%
  \bibinfo {author} {\bibfnamefont{A.}~\bibnamefont{Ibarra}}\ and\ \bibinfo
  {author} {\bibfnamefont{G.~G.}\ \bibnamefont{Ross}},\ }%
  \bibfield{journal}{%
  \Doi{10.1016/j.physletb.2004.04.037}{\bibinfo {journal} {Phys.Lett.}}\ }%
  \textbf{\bibinfo {volume} {B591}},\ \bibinfo {pages} {285} (\bibinfo {year}
  {2004}),\ \Eprint{http://arxiv.org/abs/hep-ph/0312138}{arXiv:hep-ph/0312138
  [hep-ph]}%
  \bibAnnoteFile{NoStop}{Ibarra:2003up}%
%%CITATION = HEP-PH/0312138;%%
\bibitem{Pascoli:2003rq}%
  \BibitemOpen
  \bibfield{author}{%
  \bibinfo {author} {\bibfnamefont{S.}~\bibnamefont{Pascoli}}, \bibinfo
  {author} {\bibfnamefont{S.}~\bibnamefont{Petcov}},\ and\ \bibinfo {author}
  {\bibfnamefont{C.}~\bibnamefont{Yaguna}},\ }%
  \bibfield{journal}{%
  \Doi{10.1016/S0370-2693(03)00698-1}{\bibinfo {journal} {Phys.Lett.}}\ }%
  \textbf{\bibinfo {volume} {B564}},\ \bibinfo {pages} {241} (\bibinfo {year}
  {2003}),\ \Eprint{http://arxiv.org/abs/hep-ph/0301095}{arXiv:hep-ph/0301095
  [hep-ph]}%
  \bibAnnoteFile{NoStop}{Pascoli:2003rq}%
%%CITATION = HEP-PH/0301095;%%
\bibitem{Goswami:2013lba}%
  \BibitemOpen
  \bibfield{author}{%
  \bibinfo {author} {\bibfnamefont{S.}~\bibnamefont{Goswami}}, \bibinfo
  {author} {\bibfnamefont{S.}~\bibnamefont{Khan}},\ and\ \bibinfo {author}
  {\bibfnamefont{S.}~\bibnamefont{Mishra}},\ }%
  \bibfield{journal}{%
  \Doi{10.1142/S0217751X14501140}{\bibinfo {journal} {Int.J.Mod.Phys.}}\ }%
  \textbf{\bibinfo {volume} {A29}},\ \bibinfo {pages} {1450114} (\bibinfo
  {year} {2014}),\ \Eprint{http://arxiv.org/abs/1310.1468}{arXiv:1310.1468
  [hep-ph]}%
  \bibAnnoteFile{NoStop}{Goswami:2013lba}%
%%CITATION = ARXIV:1310.1468;%%
\bibitem{Einhorn:1992um}%
  \BibitemOpen
  \bibfield{author}{%
  \bibinfo {author} {\bibfnamefont{M.}~\bibnamefont{Einhorn}}\ and\ \bibinfo
  {author} {\bibfnamefont{D.}~\bibnamefont{Jones}},\ }%
  \bibfield{journal}{%
  \Doi{10.1103/PhysRevD.46.5206}{\bibinfo {journal} {Phys.Rev.}}\ }%
  \textbf{\bibinfo {volume} {D46}},\ \bibinfo {pages} {5206} (\bibinfo {year}
  {1992})%
  \bibAnnoteFile{NoStop}{Einhorn:1992um}%
%%CITATION = PHRVA,D46,5206;%%
\bibitem{Luo:2002ey}%
  \BibitemOpen
  \bibfield{author}{%
  \bibinfo {author} {\bibfnamefont{M.-x.}\ \bibnamefont{Luo}}\ and\ \bibinfo
  {author} {\bibfnamefont{Y.}~\bibnamefont{Xiao}},\ }%
  \bibfield{journal}{%
  \Doi{10.1103/PhysRevLett.90.011601}{\bibinfo {journal} {Phys.Rev.Lett.}}\ }%
  \textbf{\bibinfo {volume} {90}},\ \bibinfo {pages} {011601} (\bibinfo {year}
  {2003}),\ \Eprint{http://arxiv.org/abs/hep-ph/0207271}{arXiv:hep-ph/0207271
  [hep-ph]}%
  \bibAnnoteFile{NoStop}{Luo:2002ey}%
%%CITATION = HEP-PH/0207271;%%
\bibitem{Machacek:1983tz}%
  \BibitemOpen
  \bibfield{author}{%
  \bibinfo {author} {\bibfnamefont{M.~E.}\ \bibnamefont{Machacek}}\ and\
  \bibinfo {author} {\bibfnamefont{M.~T.}\ \bibnamefont{Vaughn}},\ }%
  \bibfield{journal}{%
  \Doi{10.1016/0550-3213(83)90610-7}{\bibinfo {journal} {Nucl.Phys.}}\ }%
  \textbf{\bibinfo {volume} {B222}},\ \bibinfo {pages} {83} (\bibinfo {year}
  {1983})%
  \bibAnnoteFile{NoStop}{Machacek:1983tz}%
%%CITATION = NUPHA,B222,83;%%
\bibitem{Machacek:1983fi}%
  \BibitemOpen
  \bibfield{author}{%
  \bibinfo {author} {\bibfnamefont{M.~E.}\ \bibnamefont{Machacek}}\ and\
  \bibinfo {author} {\bibfnamefont{M.~T.}\ \bibnamefont{Vaughn}},\ }%
  \bibfield{journal}{%
  \Doi{10.1016/0550-3213(84)90533-9}{\bibinfo {journal} {Nucl.Phys.}}\ }%
  \textbf{\bibinfo {volume} {B236}},\ \bibinfo {pages} {221} (\bibinfo {year}
  {1984})%
  \bibAnnoteFile{NoStop}{Machacek:1983fi}%
%%CITATION = NUPHA,B236,221;%%
\bibitem{Machacek:1984zw}%
  \BibitemOpen
  \bibfield{author}{%
  \bibinfo {author} {\bibfnamefont{M.~E.}\ \bibnamefont{Machacek}}\ and\
  \bibinfo {author} {\bibfnamefont{M.~T.}\ \bibnamefont{Vaughn}},\ }%
  \bibfield{journal}{%
  \Doi{10.1016/0550-3213(85)90040-9}{\bibinfo {journal} {Nucl.Phys.}}\ }%
  \textbf{\bibinfo {volume} {B249}},\ \bibinfo {pages} {70} (\bibinfo {year}
  {1985})%
  \bibAnnoteFile{NoStop}{Machacek:1984zw}%
%%CITATION = NUPHA,B249,70;%%
\bibitem{Antusch:2002rr}%
  \BibitemOpen
  \bibfield{author}{%
  \bibinfo {author} {\bibfnamefont{S.}~\bibnamefont{Antusch}}, \bibinfo
  {author} {\bibfnamefont{J.}~\bibnamefont{Kersten}}, \bibinfo {author}
  {\bibfnamefont{M.}~\bibnamefont{Lindner}},\ and\ \bibinfo {author}
  {\bibfnamefont{M.}~\bibnamefont{Ratz}},\ }%
  \bibfield{journal}{%
  \Doi{10.1016/S0370-2693(02)01960-3}{\bibinfo {journal} {Phys.Lett.}}\ }%
  \textbf{\bibinfo {volume} {B538}},\ \bibinfo {pages} {87} (\bibinfo {year}
  {2002}),\ \Eprint{http://arxiv.org/abs/hep-ph/0203233}{arXiv:hep-ph/0203233
  [hep-ph]}%
  \bibAnnoteFile{NoStop}{Antusch:2002rr}%
%%CITATION = HEP-PH/0203233;%%
\bibitem{Mihaila:2012fm}%
  \BibitemOpen
  \bibfield{author}{%
  \bibinfo {author} {\bibfnamefont{L.~N.}\ \bibnamefont{Mihaila}}, \bibinfo
  {author} {\bibfnamefont{J.}~\bibnamefont{Salomon}},\ and\ \bibinfo {author}
  {\bibfnamefont{M.}~\bibnamefont{Steinhauser}},\ }%
  \bibfield{journal}{%
  \Doi{10.1103/PhysRevLett.108.151602}{\bibinfo {journal} {Phys.Rev.Lett.}}\ }%
  \textbf{\bibinfo {volume} {108}},\ \bibinfo {pages} {151602} (\bibinfo {year}
  {2012}),\ \Eprint{http://arxiv.org/abs/1201.5868}{arXiv:1201.5868 [hep-ph]}%
  \bibAnnoteFile{NoStop}{Mihaila:2012fm}%
%%CITATION = ARXIV:1201.5868;%%
\bibitem{Chetyrkin:2012rz}%
  \BibitemOpen
  \bibfield{author}{%
  \bibinfo {author} {\bibfnamefont{K.}~\bibnamefont{Chetyrkin}}\ and\ \bibinfo
  {author} {\bibfnamefont{M.}~\bibnamefont{Zoller}},\ }%
  \bibfield{journal}{%
  \Doi{10.1007/JHEP06(2012)033}{\bibinfo {journal} {JHEP}}\ }%
  \textbf{\bibinfo {volume} {1206}},\ \bibinfo {pages} {033} (\bibinfo {year}
  {2012}),\ \Eprint{http://arxiv.org/abs/1205.2892}{arXiv:1205.2892 [hep-ph]}%
  \bibAnnoteFile{NoStop}{Chetyrkin:2012rz}%
%%CITATION = ARXIV:1205.2892;%%
\bibitem{Melnikov:2000qh}%
  \BibitemOpen
  \bibfield{author}{%
  \bibinfo {author} {\bibfnamefont{K.}~\bibnamefont{Melnikov}}\ and\ \bibinfo
  {author} {\bibfnamefont{T.~v.}\ \bibnamefont{Ritbergen}},\ }%
  \bibfield{journal}{%
  \Doi{10.1016/S0370-2693(00)00507-4}{\bibinfo {journal} {Phys.Lett.}}\ }%
  \textbf{\bibinfo {volume} {B482}},\ \bibinfo {pages} {99} (\bibinfo {year}
  {2000}),\ \Eprint{http://arxiv.org/abs/hep-ph/9912391}{arXiv:hep-ph/9912391
  [hep-ph]}%
  \bibAnnoteFile{NoStop}{Melnikov:2000qh}%
%%CITATION = HEP-PH/9912391;%%
\bibitem{Hempfling:1994ar}%
  \BibitemOpen
  \bibfield{author}{%
  \bibinfo {author} {\bibfnamefont{R.}~\bibnamefont{Hempfling}}\ and\ \bibinfo
  {author} {\bibfnamefont{B.~A.}\ \bibnamefont{Kniehl}},\ }%
  \bibfield{journal}{%
  \Doi{10.1103/PhysRevD.51.1386}{\bibinfo {journal} {Phys.Rev.}}\ }%
  \textbf{\bibinfo {volume} {D51}},\ \bibinfo {pages} {1386} (\bibinfo {year}
  {1995}),\ \Eprint{http://arxiv.org/abs/hep-ph/9408313}{arXiv:hep-ph/9408313
  [hep-ph]}%
  \bibAnnoteFile{NoStop}{Hempfling:1994ar}%
%%CITATION = HEP-PH/9408313;%%
\bibitem{Schrempp:1996fb}%
  \BibitemOpen
  \bibfield{author}{%
  \bibinfo {author} {\bibfnamefont{B.}~\bibnamefont{Schrempp}}\ and\ \bibinfo
  {author} {\bibfnamefont{M.}~\bibnamefont{Wimmer}},\ }%
  \bibfield{journal}{%
  \Doi{10.1016/0146-6410(96)00059-2}{\bibinfo {journal}
  {Prog.Part.Nucl.Phys.}}\ }%
  \textbf{\bibinfo {volume} {37}},\ \bibinfo {pages} {1} (\bibinfo {year}
  {1996}),\ \Eprint{http://arxiv.org/abs/hep-ph/9606386}{arXiv:hep-ph/9606386
  [hep-ph]}%
  \bibAnnoteFile{NoStop}{Schrempp:1996fb}%
%%CITATION = HEP-PH/9606386;%%
\bibitem{Jegerlehner:2003py}%
  \BibitemOpen
  \bibfield{author}{%
  \bibinfo {author} {\bibfnamefont{F.}~\bibnamefont{Jegerlehner}}\ and\
  \bibinfo {author} {\bibfnamefont{M.~Y.}\ \bibnamefont{Kalmykov}},\ }%
  \bibfield{journal}{%
  \Doi{10.1016/j.nuclphysb.2003.10.012}{\bibinfo {journal} {Nucl.Phys.}}\ }%
  \textbf{\bibinfo {volume} {B676}},\ \bibinfo {pages} {365} (\bibinfo {year}
  {2004}),\ \Eprint{http://arxiv.org/abs/hep-ph/0308216}{arXiv:hep-ph/0308216
  [hep-ph]}%
  \bibAnnoteFile{NoStop}{Jegerlehner:2003py}%
%%CITATION = HEP-PH/0308216;%%
\bibitem{Sirlin:1985ux}%
  \BibitemOpen
  \bibfield{author}{%
  \bibinfo {author} {\bibfnamefont{A.}~\bibnamefont{Sirlin}}\ and\ \bibinfo
  {author} {\bibfnamefont{R.}~\bibnamefont{Zucchini}},\ }%
  \bibfield{journal}{%
  \Doi{10.1016/0550-3213(86)90096-9}{\bibinfo {journal} {Nucl.Phys.}}\ }%
  \textbf{\bibinfo {volume} {B266}},\ \bibinfo {pages} {389} (\bibinfo {year}
  {1986})%
  \bibAnnoteFile{NoStop}{Sirlin:1985ux}%
%%CITATION = NUPHA,B266,389;%%
\bibitem{Casas:1994qy}%
  \BibitemOpen
  \bibfield{author}{%
  \bibinfo {author} {\bibfnamefont{J.}~\bibnamefont{Casas}}, \bibinfo {author}
  {\bibfnamefont{J.}~\bibnamefont{Espinosa}},\ and\ \bibinfo {author}
  {\bibfnamefont{M.}~\bibnamefont{Quiros}},\ }%
  \bibfield{journal}{%
  \Doi{10.1016/0370-2693(94)01404-Z}{\bibinfo {journal} {Phys.Lett.}}\ }%
  \textbf{\bibinfo {volume} {B342}},\ \bibinfo {pages} {171} (\bibinfo {year}
  {1995}),\ \Eprint{http://arxiv.org/abs/hep-ph/9409458}{arXiv:hep-ph/9409458
  [hep-ph]}%
  \bibAnnoteFile{NoStop}{Casas:1994qy}%
%%CITATION = HEP-PH/9409458;%%
\bibitem{Casas:1996aq}%
  \BibitemOpen
  \bibfield{author}{%
  \bibinfo {author} {\bibfnamefont{J.}~\bibnamefont{Casas}}, \bibinfo {author}
  {\bibfnamefont{J.}~\bibnamefont{Espinosa}},\ and\ \bibinfo {author}
  {\bibfnamefont{M.}~\bibnamefont{Quiros}},\ }%
  \bibfield{journal}{%
  \Doi{10.1016/0370-2693(96)00682-X}{\bibinfo {journal} {Phys.Lett.}}\ }%
  \textbf{\bibinfo {volume} {B382}},\ \bibinfo {pages} {374} (\bibinfo {year}
  {1996}),\ \Eprint{http://arxiv.org/abs/hep-ph/9603227}{arXiv:hep-ph/9603227
  [hep-ph]}%
  \bibAnnoteFile{NoStop}{Casas:1996aq}%
%%CITATION = HEP-PH/9603227;%%
\bibitem{Coleman:1977py}%
  \BibitemOpen
  \bibfield{author}{%
  \bibinfo {author} {\bibfnamefont{S.~R.}\ \bibnamefont{Coleman}},\ }%
  \bibfield{journal}{%
  \Doi{10.1103/PhysRevD.15.2929, 10.1103/PhysRevD.16.1248}{\bibinfo {journal}
  {Phys.Rev.}}\ }%
  \textbf{\bibinfo {volume} {D15}},\ \bibinfo {pages} {2929} (\bibinfo {year}
  {1977})%
  \bibAnnoteFile{NoStop}{Coleman:1977py}%
%%CITATION = PHRVA,D15,2929;%%
\bibitem{Callan:1977pt}%
  \BibitemOpen
  \bibfield{author}{%
  \bibinfo {author} {\bibfnamefont{J.}~\bibnamefont{Callan},
  \bibfnamefont{Curtis~G.}}\ and\ \bibinfo {author} {\bibfnamefont{S.~R.}\
  \bibnamefont{Coleman}},\ }%
  \bibfield{journal}{%
  \Doi{10.1103/PhysRevD.16.1762}{\bibinfo {journal} {Phys.Rev.}}\ }%
  \textbf{\bibinfo {volume} {D16}},\ \bibinfo {pages} {1762} (\bibinfo {year}
  {1977})%
  \bibAnnoteFile{NoStop}{Callan:1977pt}%
%%CITATION = PHRVA,D16,1762;%%
\bibitem{Ade:2013sjv}%
  \BibitemOpen
  \bibfield{author}{%
  \bibinfo {author} {\bibfnamefont{P.}~\bibnamefont{Ade}} \emph{et~al.}
  (\bibinfo {collaboration} {Planck}),\ }%
  \bibfield{journal}{%
  \Doi{10.1051/0004-6361/201321529}{\bibinfo {journal} {Astron.Astrophys.}}\ }%
  \textbf{\bibinfo {volume} {571}},\ \bibinfo {pages} {A1} (\bibinfo {year}
  {2014}),\ \Eprint{http://arxiv.org/abs/1303.5062}{arXiv:1303.5062
  [astro-ph.CO]}%
  \bibAnnoteFile{NoStop}{Ade:2013sjv}%
%%CITATION = ARXIV:1303.5062;%%
\bibitem{Tommasini:1995ii}%
  \BibitemOpen
  \bibfield{author}{%
  \bibinfo {author} {\bibfnamefont{D.}~\bibnamefont{Tommasini}}, \bibinfo
  {author} {\bibfnamefont{G.}~\bibnamefont{Barenboim}}, \bibinfo {author}
  {\bibfnamefont{J.}~\bibnamefont{Bernabeu}},\ and\ \bibinfo {author}
  {\bibfnamefont{C.}~\bibnamefont{Jarlskog}},\ }%
  \bibfield{journal}{%
  \Doi{10.1016/0550-3213(95)00201-3}{\bibinfo {journal} {Nucl.Phys.}}\ }%
  \textbf{\bibinfo {volume} {B444}},\ \bibinfo {pages} {451} (\bibinfo {year}
  {1995}),\ \Eprint{http://arxiv.org/abs/hep-ph/9503228}{arXiv:hep-ph/9503228
  [hep-ph]}%
  \bibAnnoteFile{NoStop}{Tommasini:1995ii}%
%%CITATION = HEP-PH/9503228;%%
\bibitem{Lancaster:2011wr}%
  \BibitemOpen
  \bibfield{author}{%
  \bibinfo {author} {\bibnamefont{{Tevatron Electroweak Working Group, CDF, D0
  collaboration}}}}%
   (\bibinfo {year} {2011}),\
  \Eprint{http://arxiv.org/abs/1107.5255}{arXiv:1107.5255 [hep-ex]}%
  \bibAnnoteFile{NoStop}{Lancaster:2011wr}%
%%CITATION = ARXIV:1107.5255;%%
\bibitem{Bethke:2009jm}%
  \BibitemOpen
  \bibfield{author}{%
  \bibinfo {author} {\bibfnamefont{S.}~\bibnamefont{Bethke}},\ }%
  \bibfield{journal}{%
  \Doi{10.1140/epjc/s10052-009-1173-1}{\bibinfo {journal} {Eur.Phys.J.}}\ }%
  \textbf{\bibinfo {volume} {C64}},\ \bibinfo {pages} {689} (\bibinfo {year}
  {2009}),\ \Eprint{http://arxiv.org/abs/0908.1135}{arXiv:0908.1135 [hep-ph]}%
  \bibAnnoteFile{NoStop}{Bethke:2009jm}%
%%CITATION = ARXIV:0908.1135;%%
\bibitem{Grimus:2000vj}%
  \BibitemOpen
  \bibfield{author}{%
  \bibinfo {author} {\bibfnamefont{W.}~\bibnamefont{Grimus}}\ and\ \bibinfo
  {author} {\bibfnamefont{L.}~\bibnamefont{Lavoura}},\ }%
  \bibfield{journal}{%
  \Doi{10.1088/1126-6708/2000/11/042}{\bibinfo {journal} {JHEP}}\ }%
  \textbf{\bibinfo {volume} {0011}},\ \bibinfo {pages} {042} (\bibinfo {year}
  {2000}),\ \Eprint{http://arxiv.org/abs/hep-ph/0008179}{arXiv:hep-ph/0008179
  [hep-ph]}%
  \bibAnnoteFile{NoStop}{Grimus:2000vj}%
%%CITATION = HEP-PH/0008179;%%
\bibitem{Adam:2013mnn}%
  \BibitemOpen
  \bibfield{author}{%
  \bibinfo {author} {\bibfnamefont{J.}~\bibnamefont{Adam}} \emph{et~al.}
  (\bibinfo {collaboration} {MEG Collaboration}),\ }%
  \bibfield{journal}{%
  \Doi{10.1103/PhysRevLett.110.201801}{\bibinfo {journal} {Phys.Rev.Lett.}}\ }%
  \textbf{\bibinfo {volume} {110}},\ \bibinfo {pages} {201801} (\bibinfo {year}
  {2013}),\ \Eprint{http://arxiv.org/abs/1303.0754}{arXiv:1303.0754 [hep-ex]}%
  \bibAnnoteFile{NoStop}{Adam:2013mnn}%
%%CITATION = ARXIV:1303.0754;%%
\bibitem{Tortola:2012te}%
  \BibitemOpen
  \bibfield{author}{%
  \bibinfo {author} {\bibfnamefont{D.}~\bibnamefont{Forero}}, \bibinfo {author}
  {\bibfnamefont{M.}~\bibnamefont{Tortola}},\ and\ \bibinfo {author}
  {\bibfnamefont{J.}~\bibnamefont{Valle}},\ }%
  \bibfield{journal}{%
  \Doi{10.1103/PhysRevD.86.073012}{\bibinfo {journal} {Phys.Rev.}}\ }%
  \textbf{\bibinfo {volume} {D86}},\ \bibinfo {pages} {073012} (\bibinfo {year}
  {2012}),\ \Eprint{http://arxiv.org/abs/1205.4018}{arXiv:1205.4018 [hep-ph]}%
  \bibAnnoteFile{NoStop}{Tortola:2012te}%
%%CITATION = ARXIV:1205.4018;%%
\bibitem{Mitra:2011qr}%
  \BibitemOpen
  \bibfield{author}{%
  \bibinfo {author} {\bibfnamefont{M.}~\bibnamefont{Mitra}}, \bibinfo {author}
  {\bibfnamefont{G.}~\bibnamefont{Senjanovic}},\ and\ \bibinfo {author}
  {\bibfnamefont{F.}~\bibnamefont{Vissani}},\ }%
  \bibfield{journal}{%
  \Doi{10.1016/j.nuclphysb.2011.10.035}{\bibinfo {journal} {Nucl.Phys.}}\ }%
  \textbf{\bibinfo {volume} {B856}},\ \bibinfo {pages} {26} (\bibinfo {year}
  {2012}),\ \Eprint{http://arxiv.org/abs/1108.0004}{arXiv:1108.0004 [hep-ph]}%
  \bibAnnoteFile{NoStop}{Mitra:2011qr}%
%%CITATION = ARXIV:1108.0004;%%
\bibitem{Chakrabortty:2012mh}%
  \BibitemOpen
  \bibfield{author}{%
  \bibinfo {author} {\bibfnamefont{J.}~\bibnamefont{Chakrabortty}}, \bibinfo
  {author} {\bibfnamefont{H.~Z.}\ \bibnamefont{Devi}}, \bibinfo {author}
  {\bibfnamefont{S.}~\bibnamefont{Goswami}},\ and\ \bibinfo {author}
  {\bibfnamefont{S.}~\bibnamefont{Patra}},\ }%
  \bibfield{journal}{%
  \Doi{10.1007/JHEP08(2012)008}{\bibinfo {journal} {JHEP}}\ }%
  \textbf{\bibinfo {volume} {1208}},\ \bibinfo {pages} {008} (\bibinfo {year}
  {2012}),\ \Eprint{http://arxiv.org/abs/1204.2527}{arXiv:1204.2527 [hep-ph]}%
  \bibAnnoteFile{NoStop}{Chakrabortty:2012mh}%
%%CITATION = ARXIV:1204.2527;%%
\bibitem{Tello:2010am}%
  \BibitemOpen
  \bibfield{author}{%
  \bibinfo {author} {\bibfnamefont{V.}~\bibnamefont{Tello}}, \bibinfo {author}
  {\bibfnamefont{M.}~\bibnamefont{Nemevsek}}, \bibinfo {author}
  {\bibfnamefont{F.}~\bibnamefont{Nesti}}, \bibinfo {author}
  {\bibfnamefont{G.}~\bibnamefont{Senjanovic}},\ and\ \bibinfo {author}
  {\bibfnamefont{F.}~\bibnamefont{Vissani}},\ }%
  \bibfield{journal}{%
  \Doi{10.1103/PhysRevLett.106.151801}{\bibinfo {journal} {Phys.Rev.Lett.}}\ }%
  \textbf{\bibinfo {volume} {106}},\ \bibinfo {pages} {151801} (\bibinfo {year}
  {2011}),\ \Eprint{http://arxiv.org/abs/1011.3522}{arXiv:1011.3522 [hep-ph]}%
  \bibAnnoteFile{NoStop}{Tello:2010am}%
%%CITATION = ARXIV:1011.3522;%%
\bibitem{Alwall:2011uj}%
  \BibitemOpen
  \bibfield{author}{%
  \bibinfo {author} {\bibfnamefont{J.}~\bibnamefont{Alwall}}, \bibinfo {author}
  {\bibfnamefont{M.}~\bibnamefont{Herquet}}, \bibinfo {author}
  {\bibfnamefont{F.}~\bibnamefont{Maltoni}}, \bibinfo {author}
  {\bibfnamefont{O.}~\bibnamefont{Mattelaer}},\ and\ \bibinfo {author}
  {\bibfnamefont{T.}~\bibnamefont{Stelzer}},\ }%
  \bibfield{journal}{%
  \Doi{10.1007/JHEP06(2011)128}{\bibinfo {journal} {JHEP}}\ }%
  \textbf{\bibinfo {volume} {1106}},\ \bibinfo {pages} {128} (\bibinfo {year}
  {2011}),\ \Eprint{http://arxiv.org/abs/1106.0522}{arXiv:1106.0522 [hep-ph]}%
  \bibAnnoteFile{NoStop}{Alwall:2011uj}%
%%CITATION = ARXIV:1106.0522;%%
\bibitem{Pumplin:2002vw}%
  \BibitemOpen
  \bibfield{author}{%
  \bibinfo {author} {\bibfnamefont{J.}~\bibnamefont{Pumplin}}, \bibinfo
  {author} {\bibfnamefont{D.}~\bibnamefont{Stump}}, \bibinfo {author}
  {\bibfnamefont{J.}~\bibnamefont{Huston}}, \bibinfo {author}
  {\bibfnamefont{H.}~\bibnamefont{Lai}}, \bibinfo {author}
  {\bibfnamefont{P.~M.}\ \bibnamefont{Nadolsky}}, \emph{et~al.},\ }%
  \bibfield{journal}{%
  \Doi{10.1088/1126-6708/2002/07/012}{\bibinfo {journal} {JHEP}}\ }%
  \textbf{\bibinfo {volume} {0207}},\ \bibinfo {pages} {012} (\bibinfo {year}
  {2002}),\ \Eprint{http://arxiv.org/abs/hep-ph/0201195}{arXiv:hep-ph/0201195
  [hep-ph]}%
  \bibAnnoteFile{NoStop}{Pumplin:2002vw}%
%%CITATION = HEP-PH/0201195;%%
\bibitem{Bozzi:2007ur}%
  \BibitemOpen
  \bibfield{author}{%
  \bibinfo {author} {\bibfnamefont{G.}~\bibnamefont{Bozzi}}, \bibinfo {author}
  {\bibfnamefont{B.}~\bibnamefont{Jager}}, \bibinfo {author}
  {\bibfnamefont{C.}~\bibnamefont{Oleari}},\ and\ \bibinfo {author}
  {\bibfnamefont{D.}~\bibnamefont{Zeppenfeld}},\ }%
  \bibfield{journal}{%
  \Doi{10.1103/PhysRevD.75.073004}{\bibinfo {journal} {Phys.Rev.}}\ }%
  \textbf{\bibinfo {volume} {D75}},\ \bibinfo {pages} {073004} (\bibinfo {year}
  {2007}),\ \Eprint{http://arxiv.org/abs/hep-ph/0701105}{arXiv:hep-ph/0701105
  [hep-ph]}%
  \bibAnnoteFile{NoStop}{Bozzi:2007ur}%
%%CITATION = HEP-PH/0701105;%%
\bibitem{Christensen:2008py}%
  \BibitemOpen
  \bibfield{author}{%
  \bibinfo {author} {\bibfnamefont{N.~D.}\ \bibnamefont{Christensen}}\ and\
  \bibinfo {author} {\bibfnamefont{C.}~\bibnamefont{Duhr}},\ }%
  \bibfield{journal}{%
  \Doi{10.1016/j.cpc.2009.02.018}{\bibinfo {journal} {Comput.Phys.Commun.}}\ }%
  \textbf{\bibinfo {volume} {180}},\ \bibinfo {pages} {1614} (\bibinfo {year}
  {2009}),\ \Eprint{http://arxiv.org/abs/0806.4194}{arXiv:0806.4194 [hep-ph]}%
  \bibAnnoteFile{NoStop}{Christensen:2008py}%
%%CITATION = ARXIV:0806.4194;%%
\bibitem{Alwall:2006yp}%
  \BibitemOpen
  \bibfield{author}{%
  \bibinfo {author} {\bibfnamefont{J.}~\bibnamefont{Alwall}}, \bibinfo {author}
  {\bibfnamefont{A.}~\bibnamefont{Ballestrero}}, \bibinfo {author}
  {\bibfnamefont{P.}~\bibnamefont{Bartalini}}, \bibinfo {author}
  {\bibfnamefont{S.}~\bibnamefont{Belov}}, \bibinfo {author}
  {\bibfnamefont{E.}~\bibnamefont{Boos}}, \emph{et~al.},\ }%
  \bibfield{journal}{%
  \Doi{10.1016/j.cpc.2006.11.010}{\bibinfo {journal} {Comput.Phys.Commun.}}\ }%
  \textbf{\bibinfo {volume} {176}},\ \bibinfo {pages} {300} (\bibinfo {year}
  {2007}),\ \Eprint{http://arxiv.org/abs/hep-ph/0609017}{arXiv:hep-ph/0609017
  [hep-ph]}%
  \bibAnnoteFile{NoStop}{Alwall:2006yp}%
%%CITATION = HEP-PH/0609017;%%
\bibitem{Sjostrand:2006za}%
  \BibitemOpen
  \bibfield{author}{%
  \bibinfo {author} {\bibfnamefont{T.}~\bibnamefont{Sjostrand}}, \bibinfo
  {author} {\bibfnamefont{S.}~\bibnamefont{Mrenna}},\ and\ \bibinfo {author}
  {\bibfnamefont{P.~Z.}\ \bibnamefont{Skands}},\ }%
  \bibfield{journal}{%
  \Doi{10.1088/1126-6708/2006/05/026}{\bibinfo {journal} {JHEP}}\ }%
  \textbf{\bibinfo {volume} {0605}},\ \bibinfo {pages} {026} (\bibinfo {year}
  {2006}),\ \Eprint{http://arxiv.org/abs/hep-ph/0603175}{arXiv:hep-ph/0603175
  [hep-ph]}%
  \bibAnnoteFile{NoStop}{Sjostrand:2006za}%
%%CITATION = HEP-PH/0603175;%%
\bibitem{Mangano:2002ea}%
  \BibitemOpen
  \bibfield{author}{%
  \bibinfo {author} {\bibfnamefont{M.~L.}\ \bibnamefont{Mangano}}, \bibinfo
  {author} {\bibfnamefont{M.}~\bibnamefont{Moretti}}, \bibinfo {author}
  {\bibfnamefont{F.}~\bibnamefont{Piccinini}}, \bibinfo {author}
  {\bibfnamefont{R.}~\bibnamefont{Pittau}},\ and\ \bibinfo {author}
  {\bibfnamefont{A.~D.}\ \bibnamefont{Polosa}},\ }%
  \bibfield{journal}{%
  \Doi{10.1088/1126-6708/2003/07/001}{\bibinfo {journal} {JHEP}}\ }%
  \textbf{\bibinfo {volume} {0307}},\ \bibinfo {pages} {001} (\bibinfo {year}
  {2003}),\ \Eprint{http://arxiv.org/abs/hep-ph/0206293}{arXiv:hep-ph/0206293
  [hep-ph]}%
  \bibAnnoteFile{NoStop}{Mangano:2002ea}%
%%CITATION = HEP-PH/0206293;%%
\bibitem{Bambhaniya:2013wza}%
  \BibitemOpen
  \bibfield{author}{%
  \bibinfo {author} {\bibfnamefont{G.}~\bibnamefont{Bambhaniya}}, \bibinfo
  {author} {\bibfnamefont{J.}~\bibnamefont{Chakrabortty}}, \bibinfo {author}
  {\bibfnamefont{J.}~\bibnamefont{Gluza}}, \bibinfo {author}
  {\bibfnamefont{M.}~\bibnamefont{Kordiaczynska}},\ and\ \bibinfo {author}
  {\bibfnamefont{R.}~\bibnamefont{Szafron}}}%
   (\bibinfo {year} {2013}),\
  \Eprint{http://arxiv.org/abs/1311.4144}{arXiv:1311.4144 [hep-ph]}%
  \bibAnnoteFile{NoStop}{Bambhaniya:2013wza}%
\end{thebibliography}%

\end{document}